\documentclass[fleqn,usenatbib]{mnras}
\usepackage{newtxtext,newtxmath}
\usepackage[T1]{fontenc}
\usepackage{ae,aecompl}

\usepackage{graphicx}	% Including figure files
\usepackage{amsmath}	% Advanced maths commands
\usepackage{amssymb}	% Extra maths symbols
\usepackage[usenames,dvipsnames]{xcolor}
\usepackage{hyperref}
\usepackage[normalem]{ulem}
\usepackage{soul}

\newcommand{\sub}[1]{_{\mathrm{#1}}}
\newcommand{\msun}{M$_{\sun}$}

\def\equationautorefname~#1\null{Eq.~(#1)\null}
\def\figureautorefname~#1\null{Fig.~#1\null}
\newcommand{\appref}[1]{\hyperref[#1]{Appendix~\ref{#1}}}
\title[SMBHBs in nuclear star clusters]{Accelerated orbital decay of supermassive black hole binaries in merging nuclear star clusters}

\author[G. Ogiya et al.]{
Go Ogiya$^{1,2,3}$\thanks{E-mail: gogiya@uwaterloo.ca (GO)}, 
Oliver Hahn$^{1}$, 
Chiara M. F. Mingarelli$^{4,5}$ 
and Marta Volonteri$^{6}$ 
\\
$^{1}$Laboratoire Lagrange, Universit\'e C\^ote d'Azur, Observatoire de la C\^ote d'Azur, CNRS,\\ \quad Boulevard de l'Observatoire, CS 34229, 06304 Nice, France \\
$^{2}$Waterloo Centre for Astrophysics, University of Waterloo, Waterloo, ON N2L 3G1, Canada \\
$^{3}$Department of Physics and Astronomy, University of Waterloo, 200 University Avenue West, Waterloo, Ontario N2L 3G1, Canada \\
$^{4}$Center for Computational Astrophysics, Flatiron Institute, 162 Fifth Ave, New York, NY 10010, USA \\
$^{5}$Department of Physics, University of Connecticut, 196 Auditorium Road, U-3046, Storrs, CT 06269-3046, USA\\
$^{6}$Institut d'Astrophysique de Paris, Sorbonne Universit\'e, UPMC Univ CNRS, UMR 7095,\\ \quad 98 bis Boulevard Arago, 75014 Paris, France
}

\date{Accepted XXX. Received YYY; in original form ZZZ}

\pubyear{2019}

\begin{document}
\label{firstpage}
\pagerange{\pageref{firstpage}--\pageref{lastpage}}
\maketitle

\begin{abstract}
The coalescence of supermassive black holes (SMBHs) should generate the strongest sources of gravitational waves (GWs) in the Universe. However, the dynamics of their coalescence is the subject of much debate. In this study, we use a suite of $N$-body simulations to follow the merger of two nuclear star clusters (NSCs), each hosting a SMBH in their centre. We find that the presence of distinct star clusters around each SMBH has important consequences for the dynamical evolution of the SMBH binary: (i) The separation between the SMBHs decreases by a few orders of magnitude in the first few Myrs by the combined effects of dynamical friction and a drag force caused by tidally stripped stars. In fact, this is a significant speedup for equal mass ratio binaries, and becomes extreme for unequal mass ratios, e.g. 1:10 or 1:100, which traditional dynamical friction alone would not permit to bind. (i\hspace{-.1em}i) The subsequent binary hardening is driven by the gravitational slingshots between the SMBH binary and stars, and also depends on the mass ratio between the SMBHs. Thus, with this additional drag force, we find that all SMBHs in our suite coalesce within a Hubble time. Given that about 50\% of Milky Way sized galaxies host NSCs, our results are encouraging for upcoming GW observations with the Laser Interferometer Space Antenna -- LISA -- which will detect SMBH coalescence in the $10^4-10^7$ \msun{} mass range.
\end{abstract}


%%%%%%%%%%%%%%%%%%%%%%%%%%%%%%%%%%%%%%%%%%%%%%%%%%

%%%%%%%%%%%%%%%%% BODY OF PAPER %%%%%%%%%%%%%%%%%%

%%%%%%%%%%%%%%%%%%%%%%%%%%%%%%%%%%%%%%%%%%%%%%%%%%
%%%%%%%%%%%%%%%%%%%%%%%%%%%%%%%%%%%%%%%%%%%%%%%%%%
\section{Introduction}
\label{sec:intro}
The detection of gravitational waves (GWs) by the advanced Laser Interferometer Gravitational-Wave Observatory (aLIGO) and Virgo helped to establish a new field of astronomy \citep[and subsequent detections]{GW150914}. Thus far, high-frequency GWs have been detected from merging stellar mass black holes (BHs) and neutron stars. The binary masses of the detected events range from $\sim 3$\,\msun{} to $\sim 60$\,\msun{}, which while interesting, will be dwarfed by supermassive massive black hole (SMBH) coalescence in the centres of galaxies. In the next decade, the low-frequency inspiral of the most massive SMBH binaries (SMBHBs), $\gtrsim 10^8$\,\msun{}, is expected to be detected by Pulsar Timing Arrays \citep[PTA;][]{Mingarelli2017,Kelley2018}, while the final coalescence of SMBHs in the $10^4-10^7$ \,\msun{} range will be accessible with the upcoming Laser Interferometer Space Antenna \citep[LISA;][]{Amaro-Seoane2017}, and is the subject of our study here. 

A likely formation channel of SMBHBs is through galaxy mergers, ubiquitously observed and expected by the standard paradigm of hierarchical structure formation in the Universe. After a galaxy merger, the SMBHs are expected to experience the following three phases before emitting GWs \citep{Merritt2013}. In the first stage (pre-binary phase), dynamical friction of stars and dark matter \citep[e.g.][]{Chandrasekhar1943,Antonini2012a,Ogiya2016} as well as of the interstellar gas \citep[e.g.][]{Ostriker1999,Escala2004,Tanaka2009} play a role in depleting the SMBH's angular momentum and orbital energy with respect to the centre of the merged galaxy. The SMBHs therefore sink towards the centre of the merged galaxy, and the separation between them, $d$, decreases. When $d$ falls below the gravitational influence radius of the more massive (primary) BH, 
\begin{eqnarray}
d\sub{b} \equiv \frac{GM\sub{1}}{\sigma^2}, \label{eq:rb}
\end{eqnarray}
the SMBHs form a bound binary. Here, $G$ is the gravitational constant and $M\sub{1}$ and $\sigma$ are the mass of the primary SMBH and velocity dispersion of stars, respectively. When the merged galaxy is in a virial equilibrium state, $d\sub{b}$ roughly corresponds to the radius of a sphere enclosing a stellar mass of $2M\sub{1}$.

The SMBHB then experiences a rapid orbital decay driven by the combined effects of dynamical friction and gravitational slingshots between the SMBHB and stars (combined effect phase). While this phase lasts only for a short time, $\la 10 \, \tau$, where $\tau$ is the $N$-body or H{\'e}non time unit \citep{Henon1971,Heggie2014}, $d$ decreases by one to two orders of magnitude \citep{Milosavljevic2001,Merritt2006}. 

When the specific negative binding energy of the binary exceeds the typical specific negative binding energy of stars, $\sigma^2$, the SMBHB proceeds to the hard binary phase. This condition translates to $d$ being below the hard binary separation, i.e. 
\begin{eqnarray}
d\sub{hb} \equiv \frac{G \mu}{4 \sigma^2} = \frac{M\sub{2}}{M\sub{1}+M\sub{2}} \frac{d\sub{b}}{4}, \label{eq:ahb}
\end{eqnarray}
where $M\sub{2}$ is the mass of the second SMBH ($M\sub{2} \le M\sub{1}$) and $\mu \equiv M\sub{1}M\sub{2}/(M\sub{1}+M\sub{2})$ is the reduced mass of the SMBHB. While the exact definition of the hard binary separation depends on literature, we adopt \autoref{eq:ahb} in this paper. In this phase, the motion of the two SMBHs is almost purely Keplerian.

Even after reaching $d\sub{hb}$, stars interacting with the SMBHB can extract orbital energy and angular momentum from it, so that the orbit can in principle continue to decay, although there is some debate surrounding this issue. Indeed, if not enough SMBHB-star scattering occurs during the hard binary phase, the binary stalls before it reaches the GW-emission phase -- the infamous final parsec problem. For example, in spherical systems without gas, the orbital decay of the SMBHB stops because of a deficit of low orbital energy and angular momentum stars and dark matter to interact with the SMBHB, the so-called loss cone depletion \citep{Begelman1980,Milosavljevic2003}. A number of solutions have been proposed to the final parsec problem, e.g. the importance of a non-spherical galactic potential \citep[][]{Berczik2006,Khan2013,Vasiliev2015,Gualandris2017}, which suggest that the hardening rate could be close to what is expected in the full loss cone regime \citep{Sesana2015}. Viscous interactions in circumbinary discs \citep{Escala2005,Cuadra2009,Tagawa2015,Lupi2015} are also relevant in the case of a gas-dominated nucleus, although simulations have been finding conflicting results on the sign of the torque, i.e. whether the interaction between the binary and the gas shrinks the binary separation (negative torque), or increases it (positive torque) \citep[][and references therein]{Moody2019}. Further interactions with SMBHs from subsequent galaxy mergers have also been shown to lead to their coalescence \citep{Iwasawa2006,Tanikawa2011,Ryu2018,Bonetti2018}, mostly when high eccentricities are excited through the Kozai-Lidov mechanism \citep{Kozai1962,Lidov1962}.

Nuclear Star Clusters (NSCs) -- dense stellar systems with mass density of $\rho \ga 10^6$\,\msun{}\,${\rm pc}^{-3}$, and of order $\mathcal{O}$(pc) across \citep[e.g.][and references therein]{Sanchez-Janssen2019} -- may be among the most important factors in the evolution of SMBHs in the LISA band for GW observations. The masses of NSCs appear to correlate with the mass of their host galaxies \citep[][and references therein]{Georgiev2016,Sanchez-Janssen2019}. \cite{Sanchez-Janssen2019} showed that their presence in galaxies depends on the galaxy's stellar mass, $M\sub{gal}$, and peaks at $M\sub{gal} \approx 10^9$\,\msun{}, where up to 90\% of galaxies appear to host an NSC, while the fraction drops below 20\% at $M\sub{gal} \approx 10^7$\,\msun{} and $M\sub{gal} \approx 10^{11}$\,\msun{}. A NSC and a SMBH co-exist in the centre of many galaxies, even locally, in the centre of our Milky Way \citep{Schodel2007,Ghez2008,Gillessen2009,Genzel2010}. Assuming that all NSCs host a SMBH in their centre, about 50\% of Milky Way sized galaxies should host both a NSC and a SMBH in their centre. In addition, numerical simulations of SMBHB formation through galaxy mergers find that gas compression triggers bursts of star formation at pericentres. As a result, dense NSCs are formed and the SMBHs are embedded in them \citep{VanWassenhove2014} during the last phase of the galaxy merger.  Furthermore, recent searches for SMBHs in dwarf galaxies have successfully found them \citep{Nguyen2018}, and classic analytic estimates of the SMBHB hardening timescales suggest a more rapid evolution than expected in dwarf galaxies in the presence of NSCs due to the increased stellar densities \citep{Biava2019}.

Note that the contribution of NSCs to the orbital evolution of SMBHs in the PTA band (corresponding to $\ga 10^8$\,\msun{}) would be subdominant because the mass of NSCs is not large enough with respect to the SMBH mass. Therefore we restrict our discussion in this paper to the orbital evolution of SMBHs in the LISA band.

Here we show that tidal effects from the merging NSCs accelerate the orbital evolution timescale of SMBHs before and around the time the binary is formed. In the presence of NSCs the formation of a hard binary occurs faster, accelerating the {\em whole} process of orbital decay into the GW regime. Using a suite of $N$-body simulations, we find that the relative orbit can be further efficiently shrunk by the interactions with NSC stars at the spatial scale of $\ll$\,pc, helping the binary to overcome the final parsec problem. Therefore, NSCs appear to be an important ingredient in accelerating the coalescence of SMBHBs.

This paper is organized as follows. The role of tidally stripped stars in the orbital evolution of merging NSCs is discussed with a simple analytical model in \autoref{sec:extra_drag}. We describe the simulation setup in \autoref{sec:setup} and explore the simulation results in \autoref{sec:results}. In \autoref{sec:discussion}, we discuss implications for GW observations before summarizing the paper in \autoref{sec:summary}.

%%%%%%%%%%%%%%%%%%%%%%%%%%%%%%%%%%%%%%%%%%%%%%%%%%
%%%%%%%%%%%%%%%%%%%%%%%%%%%%%%%%%%%%%%%%%%%%%%%%%%
\section{Effective drag force by stripped stars}
\label{sec:extra_drag}
In this section, we discuss how stars which were tidally stripped from their NSC can shrink a SMBHB's orbit. Let us consider that two NSCs each hosting a SMBH are orbiting each other. Stars in the outskirts of NSCs are less bound compared to those in the centres, and hence they are more easily affected by the tidal force of the other NSC. As a result, stars in the outskirts are exchanged between the NSCs or may become unbound if their orbital energy and/or angular momentum have been changed during the tidal interaction. 

\citet[][see also \citealt{Huang1956}]{Huang1963} investigated the orbital evolution of binary systems which can exchange and/or eject mass, and found that when the ejected mass reaches a distance larger than the semi-major axis of the binary, angular momentum of the binary can be carried away and the binary orbit shrinks. While they discussed the orbital evolution of binary stars via an analytical model, it is quite general and  applicable for the cases we study. 

We begin with a brief overview of the \cite{Huang1963} model. Specific angular momentum of stars in the NSC binary system and its mass are respectively denoted as $l$ and $m$. The change in $l$ through the mass loss event is 
\begin{eqnarray}
\delta l = (l\sub{s}-l) \frac{\delta m\sub{s}}{m}, \label{eq:deltah}
\end{eqnarray}
where the subscript `s' represent quantities of stripped stars. For simplicity, we suppose $\delta m\sub{s} < 0$ and $|\delta m\sub{s}| \ll m$ and that the eccentricity, $e$, of stripped stars is not changed. The latter assumption  should be valid until the stripped stars arrive at peri- or apocentre where they can be mixed effectively and thus for about an orbital period, i.e. the mixing period of the tidally stropped stars is comparable to the NSC orbital period.  Then the specific angular momentum of each component is given as
\begin{eqnarray}
l        &=& \sqrt{Gma(1-e^2)}                \label{eq:h}  \\
l\sub{s} &=& \sqrt{Gm(a + \delta a)(1-e^2)}   \label{eq:hs} 
\end{eqnarray}
where $a$ and $a + \delta a$ are the semi-major axis of the NSC binary and the typical semi-major axis of stripped stars, respectively. Because $\delta m\sub{s} < 0$, the condition to lose specific angular momentum by tidal stripping, $\delta l < 0$, is 
\begin{eqnarray}
\delta a > 0. \label{eq:minusdh_condition}
\end{eqnarray}
The exchange of angular momentum during the NSC merger process leads to an expansion of the orbit of the stripped stars. This in turn reduces the angular momentum of the NSC binary leading to orbital decay of their central SMBHs. The tidally stripped material thus exerts a net drag force onto the binary \citep[e.g.][]{Fujii2006,Fellhauer2007,vandenBosch2018,Ogiya2019}.

\begin{figure}
\begin{center}
\includegraphics[width=0.48\textwidth]{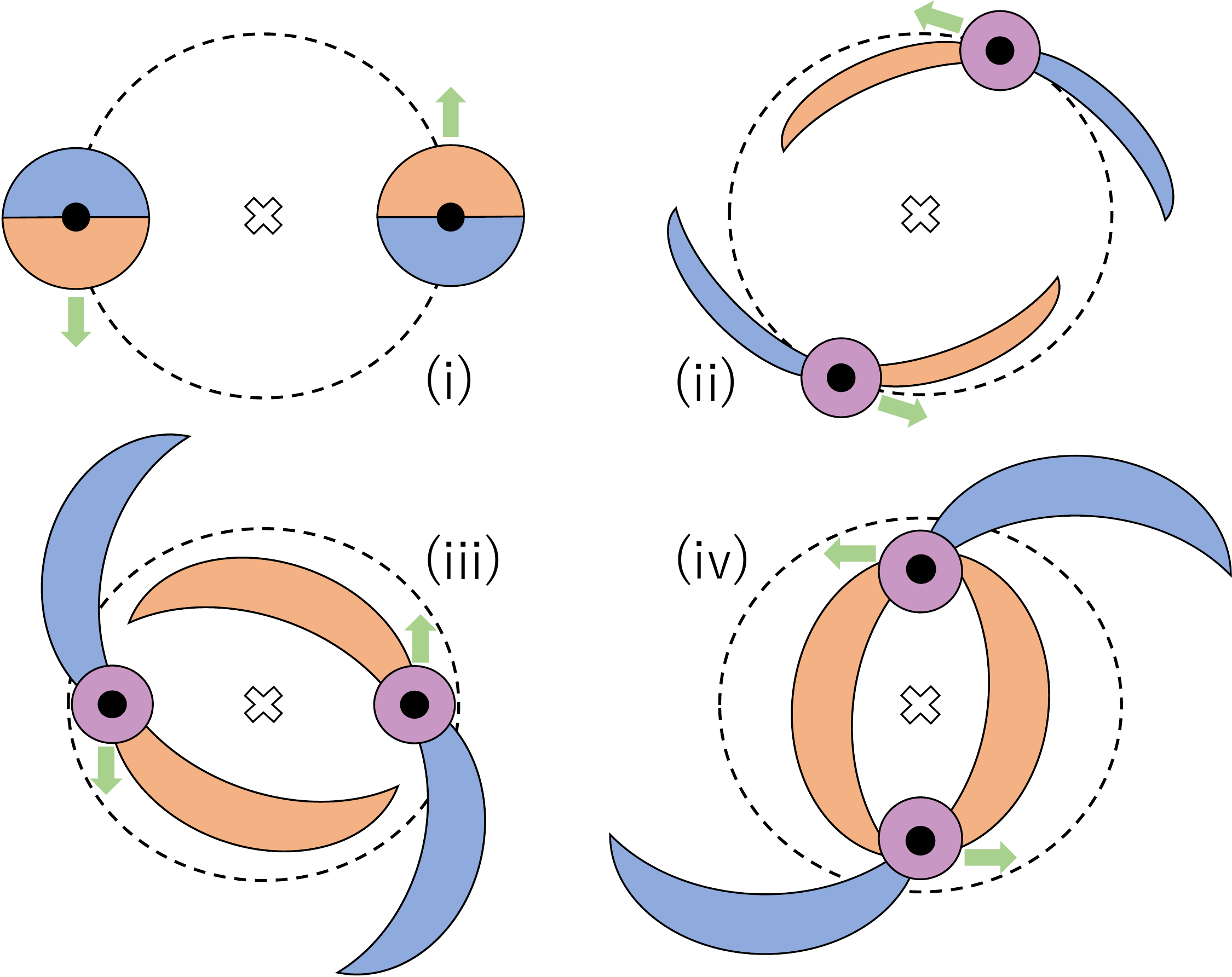}
\end{center}
\caption{
Concept of the Ouroboros Effect. Time evolution is illustrated in four panels. Initially, two NSCs each hosting a SMBH (black circle) at the centre are orbiting anticlockwise. The cross and dashed circle are the centre of mass of the entire system and initial relative orbit of the SMBHs. 
(i) Stars on the downstream (red) are decelerated by their own SMBH and stars on the upstream side (blue). As a back-reaction, stars on the upstream side are accelerated. 
(i\hspace{-.1em}i) Because of the tidal force of the other NSC, the stars on the downstream and upstream sides form a leading arm and trailing tail, respectively, while stars in the centre of the NSC surround the SMBH since they are tightly bound and resilient to the tidal force (purple). 
(i\hspace{-.1em}i\hspace{-.1em}i) The leading arm and trailing tail grow with time. Because the leading arm (trailing tail) consists of the decelerated (accelerated stars), its orbit is shrunk (expanded). 
(i\hspace{-.1em}v) The leading arm gets close to the SMBH of the other NSC and decelerates it. The two leading arms are like a pair of snakes biting each other's tail (SMBH). We describe this process in more detail in  \autoref{ssec:ouroboros}. 
\label{fig:schematic_ouroboros}}
\end{figure}
As we show in \autoref{sec:results}, the model by \cite{Huang1963} is a macroscopic description for the rapid orbital decay of SMBHs. The more microscopic description we find in \autoref{sec:results} is that the leading arm of the NSC attaches to and decelerates the companion NSC's SMBH, so that the binary NSCs appear like a pair of snakes biting each other's tail. Hereafter we refer to this type of drag force as the Ouroboros Effect, shown in \autoref{fig:schematic_ouroboros}.

%%%%%%%%%%%%%%%%%%%%%%%%%%%%%%%%%%%%%%%%%%%%%%%%%%
%%%%%%%%%%%%%%%%%%%%%%%%%%%%%%%%%%%%%%%%%%%%%%%%%%
\section{Simulation setup}
\label{sec:setup}
We perform a suite of two types of collisional $N$-body simulations to investigate the orbital evolution of SMBHs in the presence of NSCs. The first type of simulation (type-M, for ``merging'') follows mergers between two NSCs, each containing a SMBH in its centre, a situation that is expected to ensue after a major galaxy merger \citep[mass ratio $\geqslant 1:4$]{VanWassenhove2014}. In these simulations both the Ouroboros Effect and dynamical friction are at play. In the second type of simulation (type-O, for ``orbiting''), we consider a scenario where the primary SMBH is located at the centre of its NSC, and the second SMBH is orbiting this primary system without a NSC of its own. In this case only dynamical friction is at work. Type-O simulations thus represent astrophysical cases where the second galaxy has been completely disrupted (typically in minor galaxy mergers with mass ratio $<1:4$), when an SMBH returns after an ejection, e.g., through a three-body interaction or a gravitational wave kick \citep[e.g.,][]{Volonteri2005}, or when NSCs are formed through mergers between globular clusters hosting massive BHs in each \citep{Antonini2012b,Mastrobuono-Battisti2014,Arca-Sedda2018}.

%%%%%%%%%%%%%%%%%%%%%%%%%%%%%%%%%%%%%%%
\subsection{Density profile of NSCs and merger setup}
\label{ssec:ICs}
For the NSCs, we model their initial density distribution using the spherical profile by \cite{Dehnen1993},
\begin{eqnarray}
\rho(r) = \frac{(3-\gamma)M\sub{nsc,tot}}{4 \pi}\frac{r\sub{0}}{r^{\gamma}(r+r\sub{0})^{4-\gamma}}, \label{eq:dens_prof}
\end{eqnarray}
where $r$, $r\sub{0}$ and $M\sub{nsc,tot}$ represent the distance from the centre, core size, and total stellar mass of the NSC, respectively. In all simulations, we assume a centrally cored profile, $\gamma=0$, and set the core size as $r\sub{0}=1.4$\,pc which leads to an effective radius of 4\,pc, consistent with the observations of NSCs with mass of $\approx 10^7$\,\msun{} \citep{Georgiev2016}. NSCs may actually have steeper density slopes producing a higher central density. For example, the density structure of the Milky Way's NSC is modelled with $\gamma=0.5$ \citep{Chatzopoulos2015}. If the central density is higher, interactions between SMBHs and stars would be more frequent: the orbital decay rate in our simulations therefore represents a lower limit.

In the type-M simulations, each NSC has a stellar mass of $M\sub{nsc,tot}=10^7$\,\msun{}. The NSC in type-O simulations has a stellar mass of $M\sub{nsc,tot}=2 \times 10^7$\,\msun{} so that the total stellar mass is the same in both simulations. We fix the mass of the primary SMBH to be $M\sub{1} = 10^{6}$\,\msun{}, and vary that of the second, $M\sub{2}$, motivated by the scatter of BH mass in NSCs \citep{Georgiev2016}. We define the mass ratio between the SMBHs as 
\begin{eqnarray}
q \equiv M\sub{2}/M\sub{1}. \label{eq:q}
\end{eqnarray}

In the type-M simulations, each NSC consists of 65,536 equal mass stellar particles and one SMBH particle, so that the total number of particles in a simulation is 131,074. In type-O simulations, the NSC has 131,072 equal mass stellar particles and two SMBH particles are included so that the total number of particles is also 131,074. In both models, the mass of the stellar $N$-body particles is $\approx 152.6$\,\msun{}. We draw the position vector of stellar particles by rejection sampling based on the density profile. Then, a SMBH is placed at the centre of the NSC with zero velocity with respect to the cluster centre. The velocity vectors of the stellar particles are drawn as follows to ensure that the NSC is in equilibrium. Assuming that the initial velocity structure of the NSC is isotropic, we can employ the Eddington formula \citep{Eddington1916} to obtain the phase-space distribution function from the density profile. The central SMBH must of course also be taken into account to compute the gravitational potential. Then we draw for each particle an isotropic unit vector and multiply it with a velocity magnitude obtained by rejection sampling from the distribution function at the particle position. We verify that the NSC model with the central SMBH is reasonably stable in isolation (\appref{app:stability}) for time scales much longer than those relevant for the physical processes we analyse in this paper.

We denote the initial separation between two SMBHs as $d\sub{i}$\footnote{In type-M simulations, $d\sub{i}$ corresponds to the initial separation between the centres of two NSCs.}. To characterize the initial relative velocity between two SMBHs\footnote{This corresponds to the initial relative bulk velocity between two NSCs in type-M simulations.}, we introduce another parameter that characterises the angular momentum of the orbit, $\eta$, and take only the stellar mass into account. The mass of the merging systems is taken to be $M\sub{*}(d\sub{i}) \equiv \sum_{n=1}^2 M\sub{nsc}(<d\sub{i}/2) = 2M\sub{nsc}(<d\sub{i}/2)$, where $M\sub{nsc}(<r)$ is the stellar mass enclosed within $r$ from the centre of the NSC with a total mass of $10^7$\,\msun{}. The initial relative velocity, $v\sub{i}$, is evaluated as
\begin{eqnarray}
v\sub{i} = \eta \sqrt{\frac{GM\sub{*}(d\sub{i})}{d\sub{i}}}. \label{eq:vi}
\end{eqnarray}
The primary SMBH is initially set at the origin with zero-velocity and the position and velocity vectors of the second SMBH are ${\bf X} = (d\sub{i}, 0, 0)$ and ${\bf V} = (0, v\sub{i}, 0)$, respectively. Note that two SMBHs are initially at the apocentre of the relative orbit and the second SMBH initially has the same specific angular momentum with respect to the primary SMBH in the simulations with the same $d\sub{i}$ and $\eta$, i.e. the Z-component of the initial specific angular momentum vector is given as ${\bar L\sub{z}} = \eta \sqrt{GM\sub{*}(d\sub{i})d\sub{i}}$.

The setups of type-O and -M simulations are similar to those in \cite{Merritt2006} and \cite{Preto2011}, respectively. The simulations by \cite{Merritt2006} studied the orbital evolution of SMBHBs in a galactic nucleus, hosting the primary SMBH at the centre and the second SMBH is orbiting in the nucleus. Indeed, \cite{Merritt2006} showed that the timescale of orbital decay due to dynamical friction depends on the mass of the second SMBH, as expected from Chandrasekhar's theory \citep{Chandrasekhar1943}. \cite{Preto2011} studied the orbital decay of SMBHBs in the non-spherical gravitational potential field caused by a merger between two NSCs. While they varied the mass ratio between the NSC and SMBH, the two SMBHs had the same mass, i.e. $q=1.0$. Motivated by observations that indicate significant scatter in the mass of SMBHs at a fixed NSC mass scale \citep{Georgiev2016}, we vary the mass of the second SMBH, $M\sub{2}$, fixing $M\sub{1}$ as well as $M\sub{nsc,tot}$, so that type-M simulations are complementary to simulations by \cite{Preto2011}. \autoref{tab:params} provides a summary of parameters adopted in the simulations. The initial separation between SMBHs, $d\sub{i}=20$ or 50\,pc, is larger than the effective radius of the NSC model (4\,pc) and large enough to prevent the SMBHs from being bound to each other initially.

Finally, we note that we do not consider here additional possible sophistications, such as non-monochromatic stellar mass functions and associated mass segregation in the NSCs. Especially the latter might play an important role by keeping more massive stars more tightly bound to the central SMBHs.

\begin{table}
\begin{center}
\caption{Summary of the simulation parameters. Column 
(1) Type of simulation. Type-M simulates a merger between two NSCs, hosting a SMBH in each centre. In type-O, the primary SMBH is settled in the centre and the second one is initially orbiting in the NSC. 
(2) Mass ratio between two SMBHs. The mass of the primary SMBH is $10^6$\,\msun{} in all simulations. 
(3) Initial separation between SMBHs in pc. 
(4) Parameter to control the initial angular momentum. 
(5) $N$-body time unit in Myr. 
\label{tab:params}}
\begin{tabular}{ccccc}
\hline 
(1)         & (2)           & (3)           & (4)           & (5)          \\
run type    & $q$           & $d\sub{i}$    & $\eta$        & $\tau$ [Myr] \\
\hline
M           & 0.01          & 20            & 1.0           & 0.102        \\
M           & 0.1           & 20            & 0.5           & 0.084        \\
M           & 0.1           & 20            & 1.0           & 0.101        \\
M           & 0.1           & 50            & 1.0           & 0.127        \\
M           & 1.0           & 20            & 1.0           & 0.088        \\
O           & 0.01          & 20            & 1.0           & 0.054        \\
O           & 0.1           & 20            & 0.5           & 0.054        \\
O           & 0.1           & 20            & 1.0           & 0.054        \\
O           & 0.1           & 50            & 1.0           & 0.055        \\
O           & 1.0           & 20            & 1.0           & 0.058        \\
\hline
\end{tabular}
\end{center}
\end{table}

%%%%%%%%%%%%%%%%%%%%%%%%%%%%%%%%%%%%%%%
\subsection{Simulation code}
\label{ssec:code}
Both the Ouroboros Effect and dynamical friction are {\it collisionless} processes since they are caused by the change in the distribution of bulk of stars, not by encounters with single stars. However, to investigate the dynamics of SMBHs in dense NSCs, especially after SMBHs form a tightly bound hard binary ($d < d\sub{bh}$), it is important to properly handle the collisional nature of the system in order to capture the hardening through stellar scattering. 

There are difficulties in solving collisional dynamics in numerical simulations, such as the requirement of accurate time integration in close encounter events, and computational expensiveness. A well established $N$-body simulation code for collisional dynamics, {\sc Nbody6} \citep{Aarseth2003}, includes key algorithms and mathematical sophistication such as block timesteps \citep{McMillan1986, Makino1991}, splitting the total force into two parts, a slowly changing part from distant particles (regular force) and local contribution changing in a shorter timescale (irregular force), based on neighbour scheme by \cite{Ahmad1973}, and the Kustaanheimo-Stiefel (KS) regularization algorithms by \citet[][see also e.g. \citealt{Saha2009}]{Kustaanheimo1965} and by \cite{Mikkola1993}, to overcome the numerical difficulties. {\sc Nbody6} has been accelerated by parallelization, Graphic Processing Units (GPUs) and Single Instruction Multiple Data (SIMD) procedures \citep[][see also e.g. \citealt{Tanikawa2012}]{Nitadori2012}. Here we use the latest descendant, {\sc Nbody6++gpu}\footnote{https://github.com/nbodyx/Nbody6ppGPU} \citep{Wang2015}, for our calculations. 

{\sc Nbody6++gpu} has several parameters controlling the accuracy of orbit integration. The parameters determining the timesteps for the regular and irregular forces, $\eta\sub{r}$ and $\eta\sub{i}$, and for the KS regularization, $\eta\sub{u}$, are 0.005, 0.005 and 0.05, respectively. In \appref{app:force}, we show that smaller $\eta\sub{r}$, $\eta\sub{i}$ and $\eta\sub{u}$ (i.e. smaller timesteps) lead to better energy conservation (but still comparable to that in the simulation with the fiducial parameter set) while the orbital evolution of the SMBHB is almost independent of them. The timestep and distance criteria for regularization search, $dt\sub{min}$ and $r\sub{min}$, and the energy criterion to distinguish soft binaries from hard binaries, $E\sub{close}$, are initially $2 \times 10^{-6}$, $5 \times 10^{-4}$ and 1, and adjusted every 0.01\,$\tau$, based on the definition of close encounters (deflection of 90 degree). The maximum number of KS regularization pairs (star-star or SMBH-star) at the same time is a few in each simulation.

The number of neighbour particles, $N\sub{nbopt}$, determines the size of the sphere containing neighbour particles that cause the irregular force. A larger $N\sub{nbopt}$ leads to a larger neighbour sphere and is expected to yield higher force accuracy while the numerical cost increases. We adopt $N\sub{nbopt}=64$ and have a relative energy conservation of $\sim$ one percent at the end of simulations ($t=20$\,Myr). In \appref{app:force}, we show that the energy conservation and the orbital evolution of the SMBHB is almost independent of $N\sub{nbopt}$.

Collisionality of the simulated systems can still be higher than in reality because the number of stellar particles is less than that of stars in NSCs. If the average mass of stars is 1\,\msun{}, a NSC with a mass of $10^7$\,\msun{} would contain $10^7$ stars. To investigate the importance of collisionality in the orbital evolution of the SMBHs, we also perform a collisionless $N$-body simulation and find that the results of collisional and collisionless simulations agree with each other when the traditional dynamical friction and the Ouroboros Effect play a key role (\appref{app:collisionless}). We also note that the collisional simulation results are insensitive to the number of stellar particles, i.e. mass resolution (see \autoref{fig:orbit_vary_q}).

%%%%%%%%%%%%%%%%%%%%%%%%%%%%%%%%%%%%%%%%%%%%%%%%%%
%%%%%%%%%%%%%%%%%%%%%%%%%%%%%%%%%%%%%%%%%%%%%%%%%%
\section{Simulation results}
\label{sec:results}

%%%%%%%%%%%%%%%%%%%%%%%%%%%%%%%%%%%%%%%
\subsection{Ouroboros Effect}
\label{ssec:ouroboros}

\begin{figure}
\begin{center}
\includegraphics[width=0.48\textwidth]{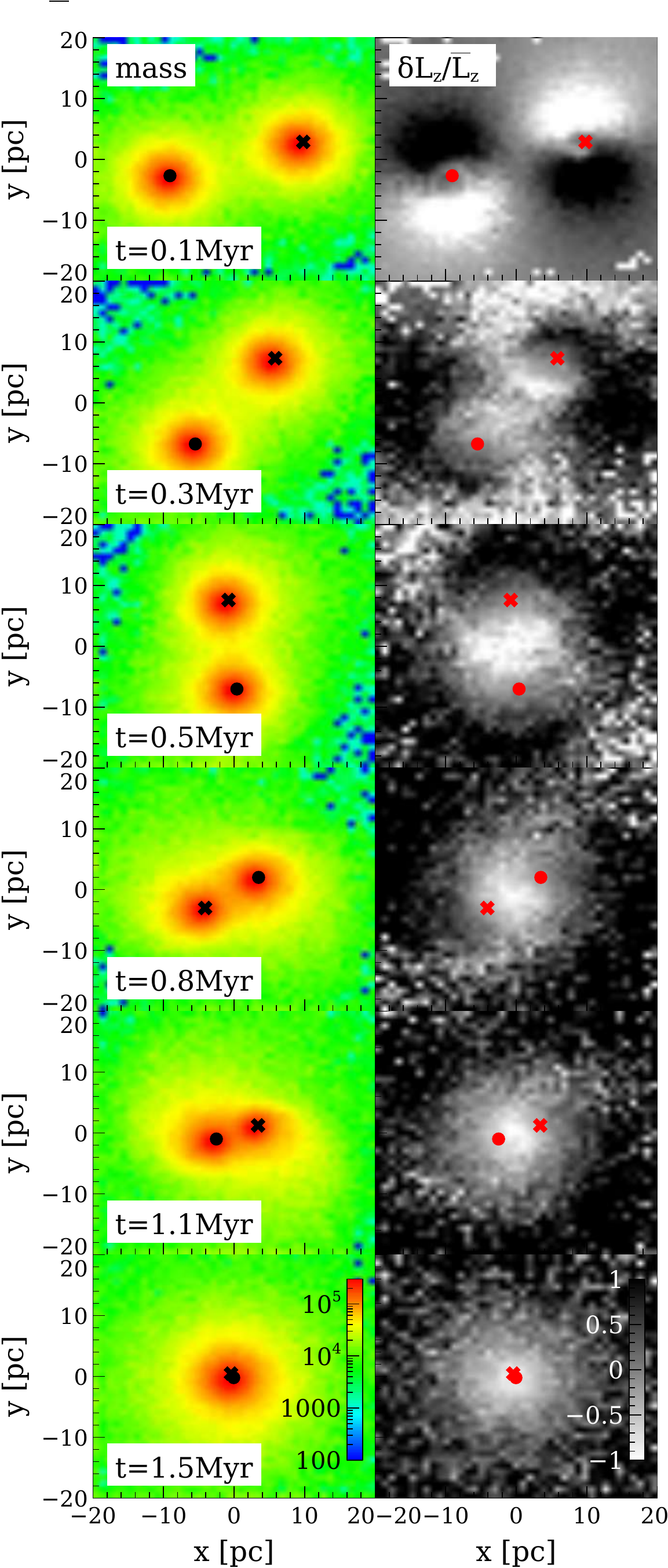}
\end{center}
\caption{
({\it Left}) Stellar mass distribution projected on the XY-plane (in \msun{}). ({\it Right}) Change in the Z-component of the angular momentum vector of each particle, $\delta L\sub{z}$, scaled by that of the initial bulk motion of the entire system, ${\bar L\sub{z}}$. Stellar particles in the range of $Z=[-10:10]$\,pc are taken into account. The origin corresponds to the centre of mass of the entire system in the type-M simulation of $q=0.1, d\sub{i}=20$\,pc and $\eta=1.0$. A circle and a cross represent primary and secondary SMBHs, respectively. Time evolution is demonstrated from top to bottom. The distance between the SMBHs is reduced to $< 1$\,pc in the first few Myrs.
\label{fig:map_mass_deltal}}
\end{figure}

Here we investigate how the Ouroboros Effect arises. The left panels of \autoref{fig:map_mass_deltal} illustrate the distribution of stellar particles in the type-M simulation of $q=0.1, d\sub{i}=20$\,pc and $\eta=1.0$. The positions of the primary and secondary SMBHs are shown as a circle and a cross. While the central parts of the two NSCs are initially separated by 20\,pc, the distance between them rapidly decreases ($\la 1$\,pc at $t=1.5$\,Myr). The timescale of orbital decay by dynamical friction is expected to be $>10$\,Myr (see \autoref{ssec:accel_decay}), so other mechanisms must be in play to drive the rapid orbital decay shown in \autoref{fig:map_mass_deltal}. 

To understand how the separation between the SMBHs decreases in such a short time, we analyse the distribution of stellar particles based on the Z-component of the angular momentum vector of each particle, $L\sub{z}$, since the initial bulk motion of the merging NSCs is anticlockwise on the XY plane with no bulk motion in the Z direction. The right panels of \autoref{fig:map_mass_deltal} demonstrate that upstream and downstream particles gain and lose $L\sub{z}$. This is because the upstream particles are pulled by the NCS core, the SMBH and central stars, while the downstream ones pulls the NSC core. This divides particles into two populations, gaining and losing angular momentum. Particles losing $L\sub{z}$ (white) fall towards the centre of the merging system, i.e. potential minimum, and particles gaining $L\sub{z}$ (black) are distributed outside.

\begin{figure}
\begin{center}
\includegraphics[width=0.4\textwidth]{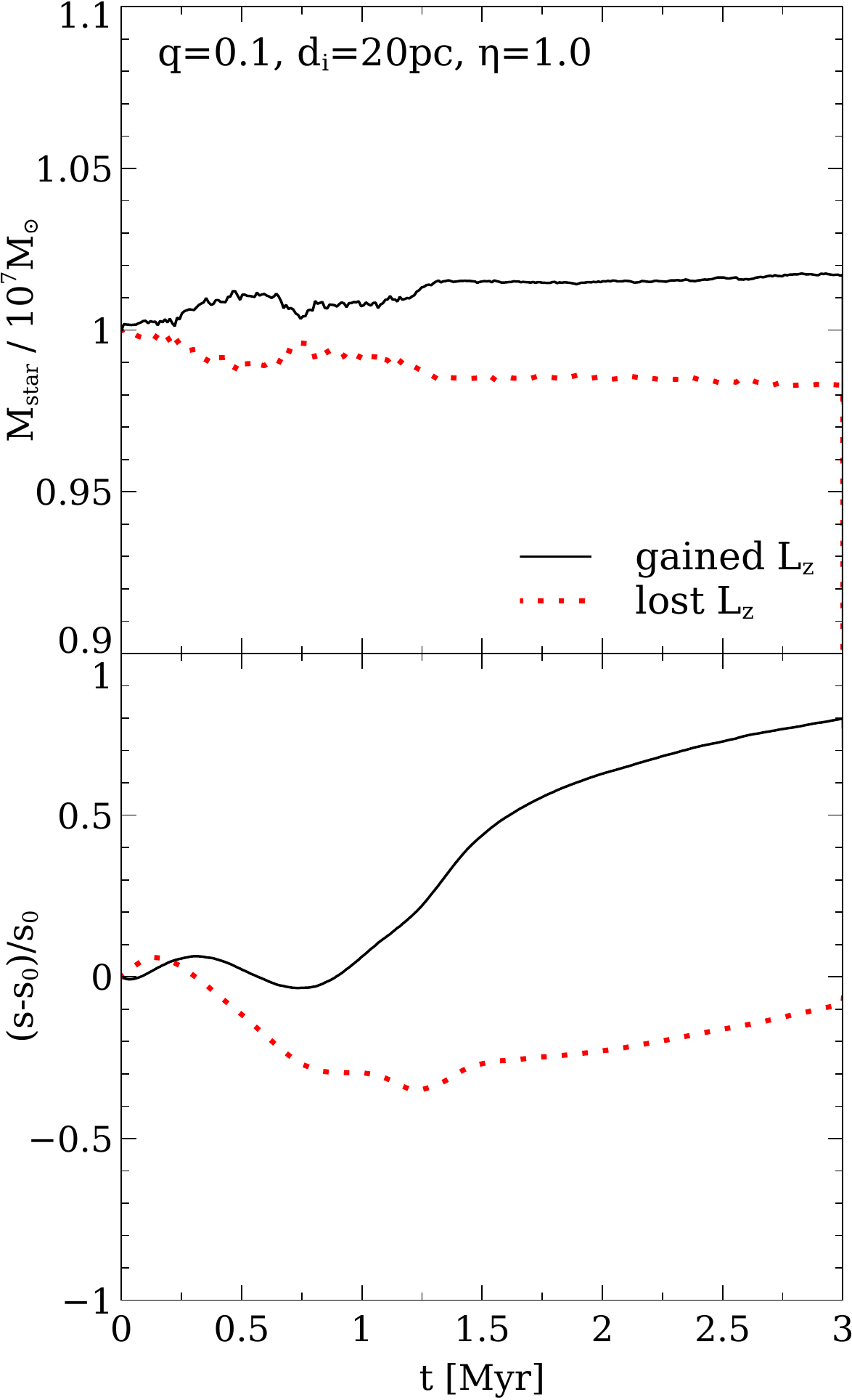}
\end{center}
\caption{
Evolution of stellar particles gaining (black solid) and losing $L\sub{z}$ (red dotted) in the type-M simulation of $q=0.1, d\sub{i}=20$\,pc and $\eta=1.0$. 
({\it Upper}) mass of each population. 
({\it Lower}) change in the distance to the centre-of-mass of the merged system, $s$. The subscript of `0' represents the initial value. The orbits of stars gaining $L\sub{z}$ expand while those of stars losing $L\sub{z}$ shrink, consistent with the model by \citet{Huang1963}. 
\label{fig:lz_gained_lost}}
\end{figure}

In \autoref{fig:lz_gained_lost}, we study the mechanism of the rapid orbital decay from a macro perspective, based on the model by \citet[][a brief review is given in \autoref{sec:extra_drag}]{Huang1963}. We track some features of stellar particles that gain (black solid) and lose (red dotted) $L\sub{z}$ during the dynamical evolution of the NSC merger in the M-type simulation with $q=0.1, d\sub{i}=20$\,pc and $\eta=1.0$. The upper panel shows that the mass of the population gaining $L\sub{z}$ is comparable to that of the population losing $L\sub{z}$ and does not significantly change with time. The lower panel presents the averaged change in the distance between the centre-of-mass of the entire system and stellar particles that belong to each population. We find that the population gaining $L\sub{z}$ moves away from the centre-of-mass of the merged system. Conversely the population losing $L\sub{z}$ moves closer to the centre-of-mass. The result is consistent with the theoretical picture by \cite{Huang1963}. Similar orbital decay process works in simulations of gaseous discs \citep{Baruteau2011}. The angular momentum of merging NSCs, each hosting a SMBHs in their centre, is extracted by the stars expanding their orbits and the orbit of the merger remnant shrinks as a back-reaction (\autoref{fig:map_mass_deltal}). Because the SMBHs are embedded in the centre of the remnant, the separation between them decreases as a consequence, facilitating the formation of the SMBHB.

\begin{figure}
\begin{center}
\includegraphics[width=0.4\textwidth]{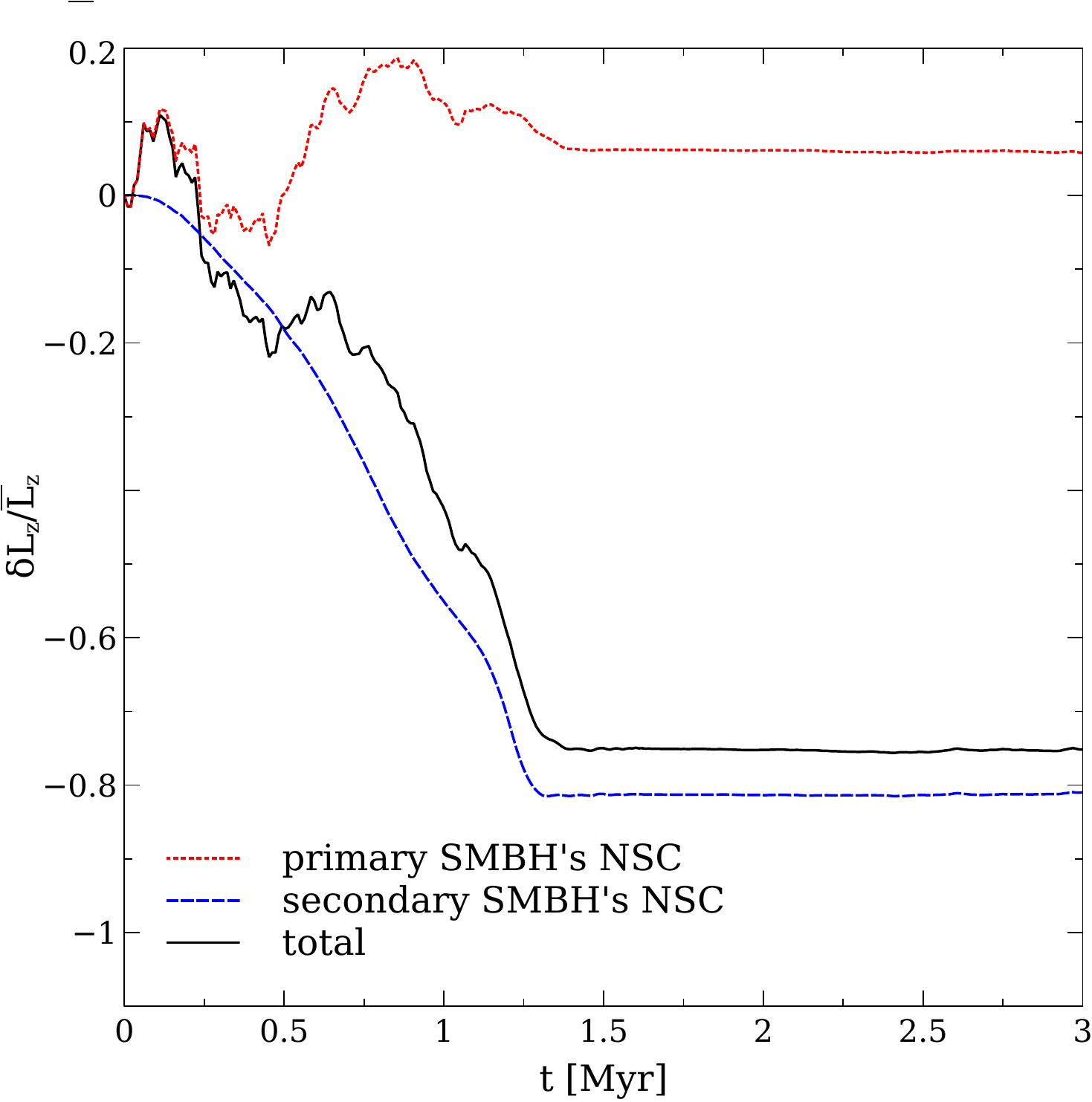}
\end{center}
\caption{
Change in $L\sub{z}$ of the primary SMBH scaled by ${\bar L\sub{z}}$. Red dotted and blue dashed lines are the contribution from stellar particles initially belonging to the NSCs hosting the primary and secondary SMBHs, respectively. Black solid line shows their sum. The primary SMBH is mainly decelerated by stars in the secondary SMBH's NSC.
\label{fig:deltal}}
\end{figure}

Which stellar particles decelerate the SMBHs? In \autoref{fig:deltal}, we show the origin of stars contributing to decrease $L\sub{z}$ of the primary SMBH. We find that the main contributors are the stellar particles initially contained in the NSC hosting the second SMBH (blue dashed). We also find that the second SMBH is mainly decelerated by stars initially belonging to the NSC hosting the primary SMBH. While the stars that initially belong to the NSC hosting the primary SMBH (red dotted) temporarily decelerate the primary SMBH, they actually accelerate it in the end. The contribution, either acceleration or deceleration, may depend on the configuration of the merger, e.g. orbit, BH mass, however a more detailed study is needed to draw a concrete conclusion. 

\begin{figure}
\begin{center}
\includegraphics[width=0.45\textwidth]{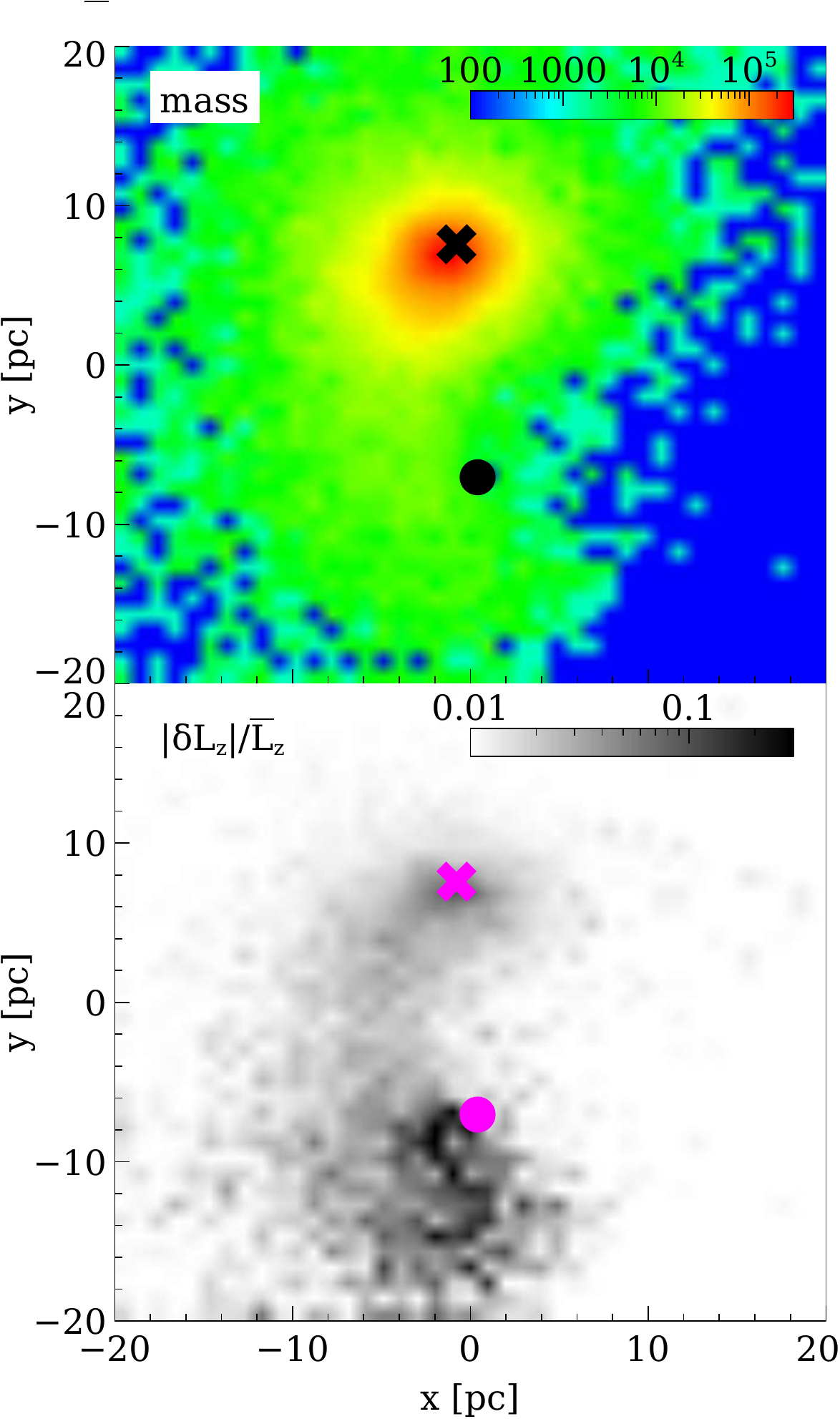}
\end{center}
\caption{
Distribution of stellar particles, that initially belong to the NSC hosting the secondary SMBH and reduce $L\sub{z}$ of the primary SMBH, projected on the XY-plane. The contribution in changing $L\sub{z}$ of the primary SMBH by the stellar particles is estimated by integrating 51 snapshots up to $t=0.5$\,Myr with the fixed time interval of $\Delta t = 0.01$\,Myr. Stellar particles in the range of $Z=[-10:10]$\,pc at $t=0.5$\,Myr are taken into account. The origin corresponds to the centre of mass of the entire system in the type-M simulation of $q=0.1, d\sub{i}=20$\,pc and $\eta=1.0$. A circle and a cross represent primary and secondary SMBHs, respectively. 
({\it Upper}) stellar mass distribution. The colour bar presents stellar mass contained in each pixel (in \msun{}). 
({\it Lower}) contribution in reducing $L\sub{z}$ of the primary SMBH, scaled by ${\bar L\sub{z}}$. 
The angular momentum of the primary SMBH is significantly reduced by the leading arm which consists of stars that initially belong to the NSC of the secondary SMBH.
\label{fig:map_neg}}
\end{figure}

Motivated by \autoref{fig:deltal}, we study the contribution of stellar particles initially belonging to the NSC of the secondary SMBH in decelerating the primary SMBH in \autoref{fig:map_neg}. The upper panel illustrates the distribution of stellar particles reducing $L\sub{z}$ of the primary SMBH and shows that the leading arm of the NSC of the second SMBH is located close to the primary SMBH (black circle). The leading arm consists of particles initially on the downstream side. Looking at the right panels of \autoref{fig:map_mass_deltal}, they are decelerated by their own NSC core, i.e. secondary SMBH and central stars, at the beginning of the merger event and fall to the potential minimum of the entire system. The lower panel shows that the primary SMBH is decelerated by these stellar particles and its $L\sub{z}$ is reduced. While \autoref{fig:map_neg} presents only the deceleration of the primary SMBH by stars initially belonging to the NSC of the secondary SMBH, we also find that stars initially belonging to the NSC of the primary SMBH decelerate the secondary SMBH in the same way.

We have shown that the Ouroboros Effect plays a key role in driving the rapid orbital decay of the SMBHs in merging NSCs. The origin of the Ouroboros Effect is summarized as follows: (i) At the beginning of a merger event between two NSCs, stars on the downstream (upstream) side are decelerated (accelerated) by the central part of their NSC, including the SMBH, and fall towards (move apart from) the centre of the entire system. (i\hspace{-.1em}i) Then, the downstream stars get close to the SMBH embedded in the other NSC and decelerate it. The timescale of the orbital decay driven by the Ouroboros Effect would be comparable to the orbital period of the NSC merger since it is triggered by the merger of the NSCs. The orbital decay timescale by dynamical friction is $\mathcal{O}(\tau\sub{orb}M\sub{nsc}/M\sub{2})$ where $\tau\sub{orb}$ is the orbital timescale of the second SMBH, $M\sub{2}$, in the NSC with a mass of $M\sub{nsc}$. Therefore the Ouroboros Effect would be more important when the second SMBH is less massive. We verify this expectation in \autoref{ssec:accel_decay}.

%%%%%%%%%%%%%%%%%%%%%%%%%%%%%%%%%%%%%%%
\subsection{Accelerated orbital decay due to the Ouroboros Effect}
\label{ssec:accel_decay}

\begin{figure*}
\begin{center}
\includegraphics[width=0.95\textwidth]{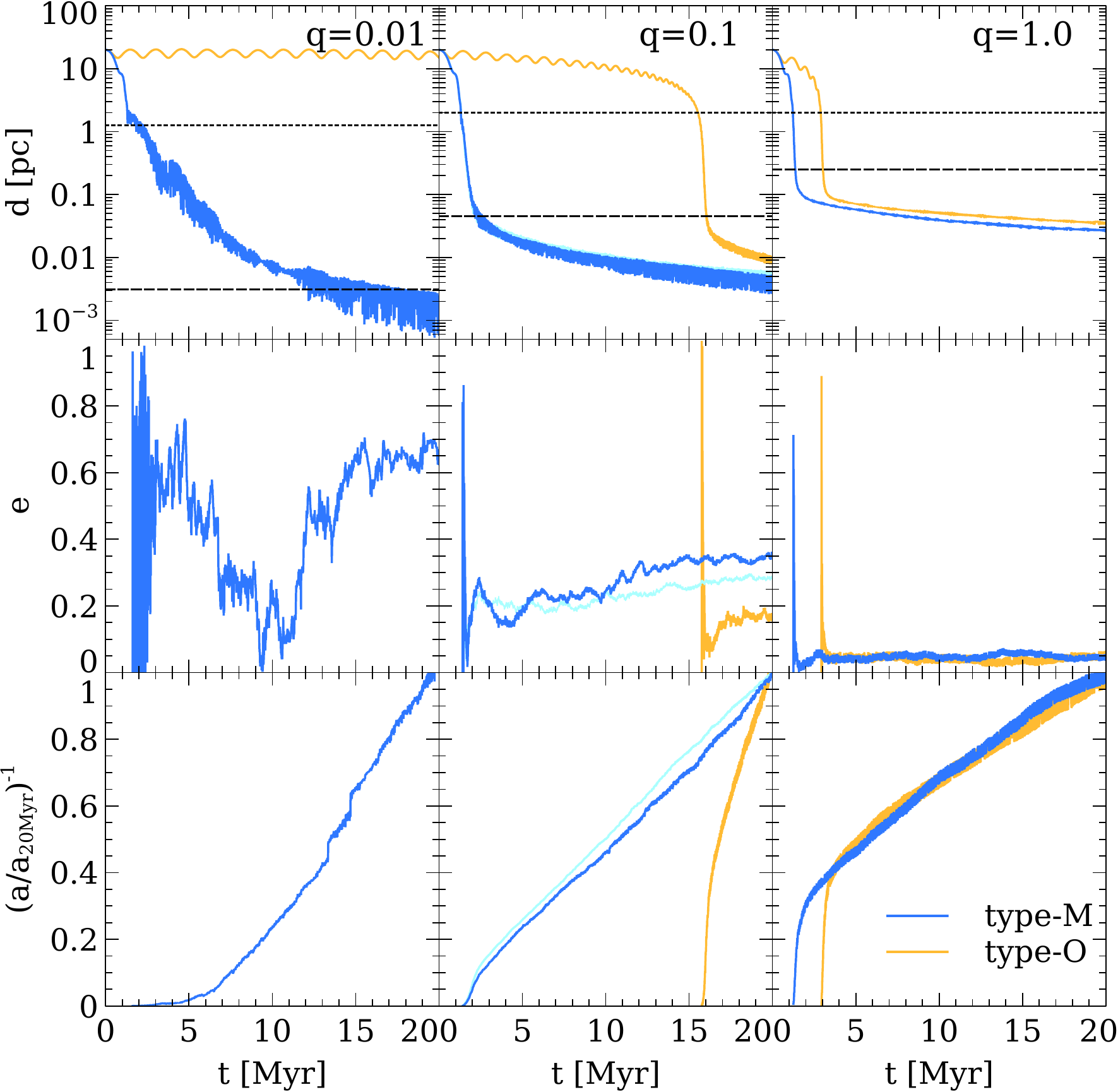}
\end{center}
\caption{
Evolution of the relative orbit of the SMBHs. Left, centre and right panels present the simulation results of $q=0.01, 0.1$ and 1.0 with the orbital parameters of $d\sub{i}=20$\,pc and $\eta=1.0$. Blue and orange lines represent the results from type-M and -O simulations.
({\it Top}) Separation between two SMBHs. Dotted and dashed horizontal lines are the gravitational influence radius of the primary SMBH, $d\sub{b}$, and hard binary separation, $d\sub{hb}$, estimated with the mass profiles of the merged system. The mass profiles are derived by stacking and averaging the snapshots in the type-M simulations. 
({\it Middle}) Eccentricity of the orbit of the SMBHB, $e$. 
({\it Bottom}) Inverse semi-major axis of the SMBHB, $1/a$, normalized by that at $t=20$\,Myr. In computing $e$ and $a$, only the two SMBHs are taken into account, i.e. stellar particles are neglected.
Cyan lines in the central panels show the results from the M-type simulation with a double number of stellar particles (for the same total stellar mass). 
The Ouroborus Effect, active in the type-M simulations but not in the type-O simulations, is responsible for the much faster orbital decay and formation of a hard binary. 
\label{fig:orbit_vary_q}}
\end{figure*}
We next study in detail how the Ouroboros Effect accelerates the orbital decay of SMBHs embedded in the centres of merging NSCs. In the top panels of \autoref{fig:orbit_vary_q}, we present the separation between the SMBHs as a function of time, $d(t)$. The orbital evolution depends on the simulation type (M or O) as well as the mass ratio between the SMBHs, $q$. It is clear that the merger of the NSCs accelerates the orbital decay, especially in the cases of low $q$ in which classical dynamical friction works inefficiently \citep{Chandrasekhar1943}. The middle and bottom panels show the eccentricity and semi-major axis evolution of the SMBHB in each simulation, taking only the two SMBHs into account for computing eccentricity, $e$, and semi-major axis, $a$, i.e. neglecting the gravity of the stellar particles. Because of this assumption, there are a few caveats regarding the evolution of the eccentricity and semi-major axis shown in the middle and bottom panels of \autoref{fig:orbit_vary_q} before the SMBHs form a hard binary, i.e. the time while the coloured lines are above the horizontal dashed lines in the top panels. Note that $e$ and $a$ are in fact not defined before the SMBHs are brought close enough. In particular, before the formation of a hard binary, the stellar potential, which we neglect in the definition of $e$ and $a$, contributes to the orbit. Once the hard binary is formed, the stellar potential can be assumed to be constant on the scales of the binary. 
The semi-major axis depends on both the orbital energy and angular momentum of the SMBHB and its smooth evolution (see bottom panels) indicates that the orbital energy and angular momentum evolve smoothly, too. This is an indication that the binary's evolution is governed by the cumulative interactions between SMBHs and stars, not by a single (perhaps artificial) violent interaction. The latter is unexpected since stars are much less massive than SMBHs and the amount of energy and angular momentum a single star can expel from the SMBHB is limited. Therefore the binary's dynamical evolution is properly resolved in the simulations. 
Cyan lines in the central panels show that these simulation results are insensitive to the number of stellar particles. 

The result that the orbital decay in our simulations is insensitive to the number of stellar particles, $N\sub{*}$, is qualitatively consistent with previous work studying the orbital decay of SMBHBs in merging galaxies or galactic nuclei \citep[][but see also \citealt{Vasiliev2015}]{Preto2011,Gualandris2017}. In spherical systems, the angular momentum and orbital energy of each star are conserved and two-body relaxation is the only mechanism to supply stars to the SMBHB after they are ejected through three-body interactions (re-filling of the loss cone). Since the timescale of two-body relaxation depends on $N\sub{*}$, the orbital decay of SMBHBs is sensitive to $N\sub{*}$ in the simulations \citep{Makino2004}. On the other hand, systems formed through mergers are not spherical \citep[e.g.][]{Preto2011,Khan2015,Khan2018b} and the loss cone is efficiently re-filled on a timescale shorter than the two-body relaxation time and that is independent of $N\sub{*}$ \citep[see e.g.][for analytical discussions]{Yu2002}. Therefore it is unsurprising that the orbital decay of SMBHBs in such systems is insensitive to $N\sub{*}$.

In the type-M simulations, the separation between the SMBHs decreases by a factor of a few orders of magnitude in the first Myrs with this efficiency depending on $q$. In the cases of $q=0.1$ and 1.0, the rapid decay driven by the Ouroboros Effect stops when $d$ drops below the hard binary separation, $d\sub{hb}$ (horizontal dashed line). The Ouroboros effect thus allows the system to bypass the pre-binary and combined effect phases and directly enter the hard binary phase. The evolutionary track to the hard binary phase in the case of $q=0.01$ is different from the others. The rapid orbital decay driven by the Ouroboros Effect stops when $d$ drops below the influence radius of the primary SMBH, $d\sub{b}$ (horizontal dotted line) that corresponds to the time to form a bound binary and enter the combined effect phase. The large difference in the masses of the SMBHs ($10^4$ and $10^6$ \,\msun{}) leads to the disruption of the central part of the NSC hosting the secondary SMBH because (i) the stars are less bound compared to those in the NSC of the primary SMBH; and (i\hspace{-.1em}i) the tidal force of the NSC that contains the primary SMBH is stronger. After the disruption, the secondary SMBH is orbiting in the stellar density field of the merged system -- a situation comparable to the set-up of the type-O simulations, and the Ouroboros Effect cannot work efficiently. The combined effect of dynamical friction and three-body interactions of the SMBHs and stars bring the SMBHB more slowly to the hard binary phase, as shown in previous studies \citep{Milosavljevic2001,Merritt2006}. While the orbital decay is less efficient, the separation $d$ still decreases by about two and a half orders of magnitude in $\sim 10$\,Myr.

After entering the hard binary phase, the orbital decay is less efficient in the type-M simulations with higher $q$. This is simply because at fixed specific angular momentum, the larger the SMBH masses the larger the absolute energy and angular momentum of the binary, and the more energy and angular momentum have to be removed from the SMBHs by the stars. Since stars increase their velocity as a back-reaction of the SMBHB orbital decay, they get ejected as the SMBHB shrinks. Eventually the stars interacting with the SMBHB dwindle because a larger stellar mass is expelled from the centre. This leads to a lower central density of the merged system and lower efficiency of orbital decay in the simulations with larger SMBH masses. This process is generally referred to as core scouring and it is the mechanism advocated for creating shallow stellar density profiles, viz. cores, in large elliptical galaxies \citep[cf.][and references therein]{Faber1997,Merritt2006,Thomas2016,Rantala2018}.

In the type-M simulation with $q=0.01$, the SMBHB orbit becomes more circular (i.e. $e$ decreases) during the combined effect phase (at $t \la 10$\,Myr). This corresponds to orbit circularization by dynamical friction. Note that it is also possible to keep or even increase $e$ with dynamical friction, depending on the density and velocity structure of the system \citep{Tsuchiya2000}. In the hard binary phase, $e$ gradually increases with time as predicted by the theoretical model for this phase \citep[e.g.][]{Sesana2015} and the resultant $e$ depends on $q$. A dedicated study with longer integration time would be needed to make more concrete conclusions regarding the $e$ evolution.

%%%%%%%%%%%%%%%%%%%%%%%%%%%%%%%%%%%%%%%%%%%%%%%%%%
\subsection{Dependence of orbital decay times on orbital parameters}
\label{ssec:orb_params}

\begin{figure}
\begin{center}
\includegraphics[width=0.45\textwidth]{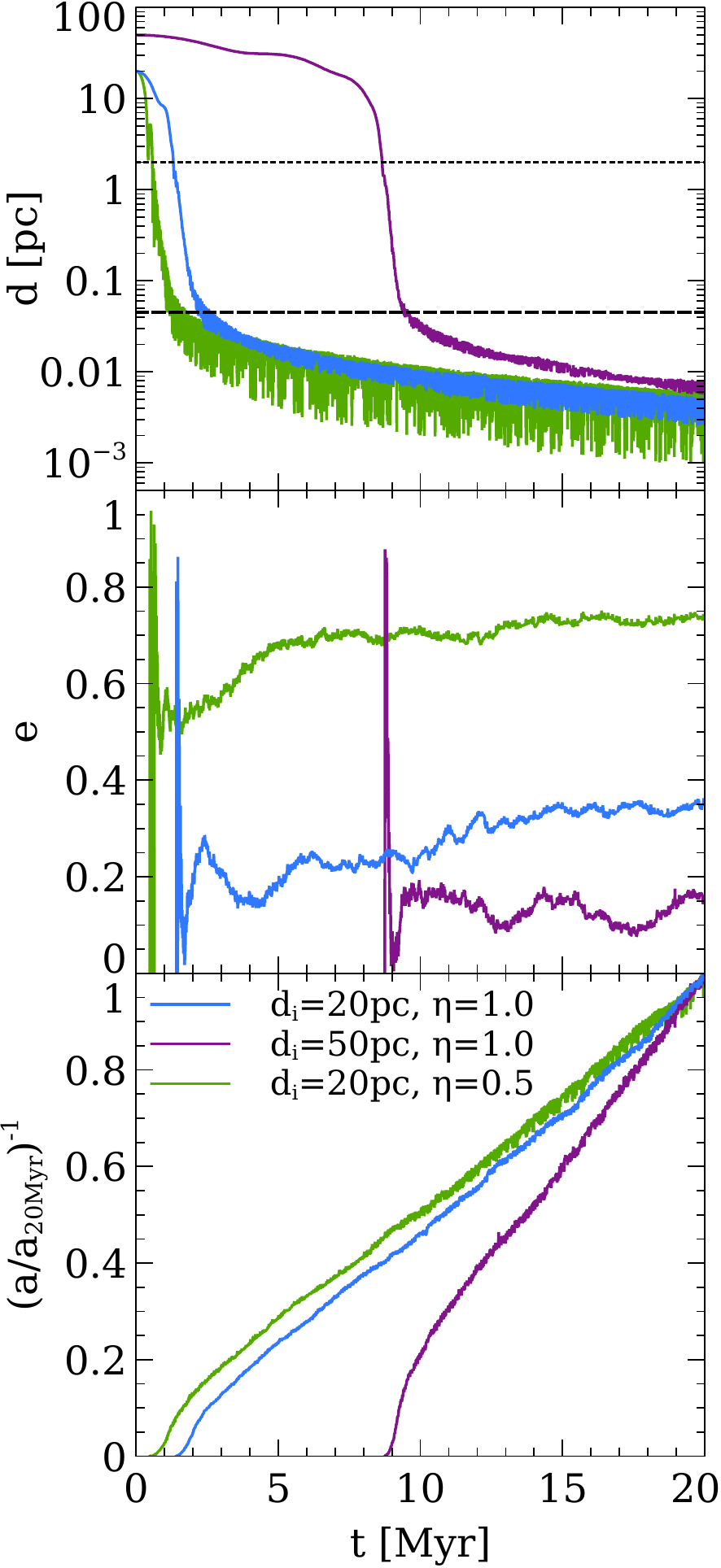}
\end{center}
\caption{
Evolution of the relative orbit of the SMBHs in the M-type simulations of $q=0.1$, varying the two orbital parameters. The adopted orbital parameters are indicated in the legend. 
({\it Top}) Separation between two SMBHs. Dotted and dashed horizontal lines are the gravitational influence radius of the primary SMBH, $d\sub{b}$, and hard binary separation, $d\sub{hb}$, estimated with the mass profiles of the merged system. The mass profiles are derived by stacking and averaging the snapshots in the type-M simulations. The estimated $d\sub{b}$ and $d\sub{hb}$ is almost independent of the orbital parameters. 
({\it Middle}) Eccentricity of the orbit of the SMBHB, $e$. 
({\it Bottom}) Inverse semi-major axis of the SMBHB, $1/a$, normalized by that at $t=20$\,Myr. In computing $e$ and $a$, only the two SMBHs are taken into account, i.e. stellar particles are neglected. 
The Ouroborus Effect accelerates the orbital decay even for significantly different orbital parameters.
\label{fig:orbit_vary_orb}}
\end{figure}
To study the dependence of the orbital decay of the SMBHs on the initial merger orbit, we vary the orbital parameters, $d\sub{i}$ and $\eta$, while fixing $q$ between the two SMBHs, and focusing on type-M models. In the simulations presented in \autoref{ssec:accel_decay}, the orbital parameters are fixed while $q$ and the configuration of the simulations are varied. Therefore the simulations in this subsection (type-M simulations of $q=0.1$ in \autoref{tab:params}) are complementary to them. 

In \autoref{fig:orbit_vary_orb}, we show the results from type-M simulations varying the orbital parameters, the initial separation between the SMBHs, $d\sub{i}$, and the parameter controlling the initial angular momentum of the merger orbit, $\eta$, while fixing $q=0.1$. The top panel shows that the time to achieve the hard binary phase strongly depends on the orbital parameters. When the merging orbit has a smaller orbital energy (viz. smaller $d\sub{i}$) or smaller angular momentum (viz. smaller $\eta$), the SMBHB enters the hard binary phase in a shorter time, since the orbital energy and angular momentum to be lost are smaller. We also found that in the type-O simulations with identical orbital parameters, the SMBHs take longer to enter the hard binary phase, $\sim 5$ and $> 20$\,Myr in the cases of $d\sub{i}=20$\,pc and $\eta=0.5$ and $d\sub{i}=50$\,pc and $\eta=1.0$, respectively (results are not shown in the figure), meaning that the Ouroborus Effect accelerates the orbital decay in all simulations in \autoref{fig:orbit_vary_orb}. The eccentricity evolution (middle panel) depends on the orbital parameters, especially $\eta$ that controls the initial angular momentum of the merging orbit. The SMBHB can have a higher eccentricity when the initial merging orbit is already more eccentric (i.e. smaller $\eta$). The semi-major axis smoothly decreases (bottom panels) and the SMBHB orbital evolution is properly resolved.

%%%%%%%%%%%%%%%%%%%%%%%%%%%%%%%%%%%%%%%%%%%%%%%%%%
%%%%%%%%%%%%%%%%%%%%%%%%%%%%%%%%%%%%%%%%%%%%%%%%%%
\section{Timescales of coalescence for SMBHs in merging NSCs}
\label{sec:discussion}

\begin{figure}
\begin{center}
\includegraphics[width=0.45\textwidth]{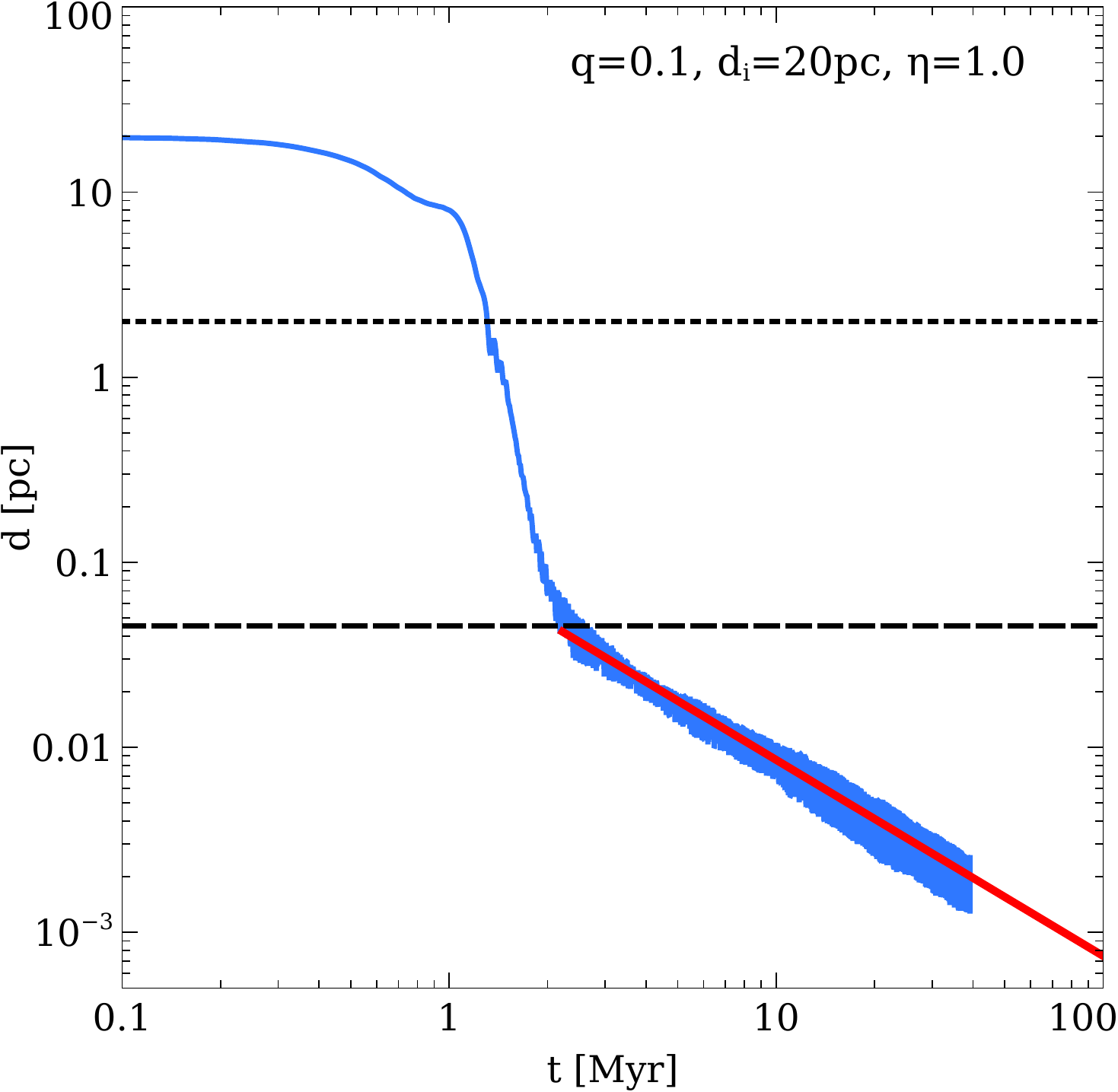}
\end{center}
\caption{
Evolution of the relative orbit of the SMBHs in the type-M simulation of $q=0.1, d\sub{i}=20$\,pc and $\eta=1.0$ (blue) on a log-log scale (the simulation is extended to $t \approx 40$\,Myr). Dotted and dashed horizontal lines represent $d\sub{b}$ and $d\sub{hb}$, respectively. In the hard binary phase, the orbital decay of the SMBHB is modelled with a power-law and the fitting result is shown as a red solid line.
}
\label{fig:orbit_loglog}
\end{figure}

Here we estimate the timescale of a SMBHB coalescence based on our simulation results. Plotting the orbital evolution of the SMBHB on a log-log scale, we find an interesting feature after the SMBHB enters the hard binary phase. As depicted in \autoref{fig:orbit_loglog}, the SMBHB orbit continues to shrink in a single power-law fashion. The power-law orbital decay is found in all type-M simulations performed in this study. We fit it with a single power-law function after the separation between the SMBH drops below $d\sub{hb}$ for the first time. The fitting parameters are derived using the least squares method.

By extrapolating the fitting results, we can discuss the timescales of SMBH coalescence all the way through to the final GW emission phase. Previous studies have developed a theoretical framework for the dynamical evolution of SMBHBs in the hard binary phase \cite[e.g.][]{Quinlan1996,Yu2002,Sesana2006,Sesana2008} and discussed interesting astrophysical phenomena, including GW emission and hypervelocity stars originated by SMBHBs \citep{Yu2003,Sesana2007}. They found that the decay of SMBHBs in the hard binary phase is described as
\begin{eqnarray}
\frac{d}{dt} \biggl ( \frac{1}{a} \biggr ) = \frac{G H \rho}{\sigma} 
\end{eqnarray}
where $a$, $\rho$ and $\sigma$ are the semi-major axis of the SMBHB, mass density and velocity dispersion of stars, respectively. The dimensionless parameter, $H$, is referred to as the binary hardening rate and depends on $a,e,q$ and the density structure of background stars. \cite{Sesana2010} showed the power-law decay of the SMBHB orbit when $H$ is independent of $a$. In our simulations the stellar density of the merged system at small radii is higher than \cite{Sesana2010} assumed, and the power-law slope of the orbital decay at $0.1 \la d/d\sub{hb} < 1$ is explained by $H \propto \rho^{-1/2}$ (i.e. lower $H$ at smaller $a$). Assuming a constant $H$, the orbital decay would be faster than we find.

Let us thus finally estimate the timescale for SMBH coalescence, $t\sub{coa}$, in type-M simulations. After entering the hard binary phase, we suppose that initially the decay is driven by stellar hardening, with a power-law decay fit from the simulation down to the scales below which GW emission dominates, for which we adopt the analytical expressions in \cite{Peters1964}, fixing the eccentricity of the SMBHB, $e'$. The transition between the two regimes occurs at the semi-major axis $a'=d(t\sub{pow})$ where the sum of $t\sub{pow}$ and $t\sub{GW}$ is minimized (see \autoref{tab:tcoalesce} for their definitions).

\begin{table*}
\begin{center}
\caption{Expected time of SMBH coalescence in the merged NSCs (i.e. type-M simulations). 
Description of each column: 
(1) Simulation parameters. In the `HR' run, the number of particles is doubled compared to the fiducial one. 
(2) Assumed eccentricity. 
(3) Semi-major axis of the SMBHB to have the minimum coalescence time.
(4) Time to drop to $a'$ by the power-law stellar hardening. 
(5) Time to lose orbital energy of the SMBHB by GW emission. 
(6) SMBH coalescence time measuring from the beginning of the NSC merger, i.e. $t\sub{coa} \equiv t\sub{pow}+t\sub{GW}$.  
\label{tab:tcoalesce}}
\begin{tabular}{cccccc}
\hline 
(1)                      & (2)  & (3)                   & (4)                & (5)                & (6)                \\
$[ q, d\sub{i}, \eta ]$  & $e'$ & $a'$ [pc]             & $t\sub{pow}$ [Myr] &  $t\sub{GW}$ [Myr] & $t\sub{coa}$ [Myr] \\
\hline
$[0.01, 20, 1.0]$        & 0.7  &  $2.3 \times 10^{-4}$ & 51.4               & 6.1                & 57.6               \\
                         & 0.0  &  $1.1 \times 10^{-4}$ & 73.2               & 8.8                & 82.0               \\
$[0.1,  20, 1.0]$        & 0.3  &  $3.6 \times 10^{-4}$ & 201.4              & 47.5               & 248.9              \\
                         & 0.0  &  $3.2 \times 10^{-4}$ & 224.7              & 53.0               & 277.8              \\
$[0.1,  20, 1.0]$\,(HR)  & 0.3  &  $3.8 \times 10^{-4}$ & 254.2              &   64.3               & 318.5              \\
                         & 0.0  &  $3.4 \times 10^{-4}$ & 285.5              &   72.2               & 357.6             \\ 
$[1.0,  20, 1.0]$        & 0.05 &  $1.5 \times 10^{-3}$ & 3585.0             & 1627.4             & 5212.4             \\
                         & 0.0  &  $1.5 \times 10^{-3}$ & 3602.9             & 1636.2             & 5239.1             \\
$[0.1,  50, 1.0]$        & 0.15 &  $2.2 \times 10^{-4}$ & 93.1               & 10.8               & 103.9              \\
                         & 0.0  &  $2.1 \times 10^{-4}$ & 94.5               & 10.9               & 105.5              \\
$[0.1,  20, 0.5]$        & 0.75 &  $7.7 \times 10^{-4}$ & 131.1              & 37.2               & 168.3              \\
                         & 0.0  &  $3.6 \times 10^{-4}$ & 312.5              & 88.6               & 401.1              \\
\hline
\end{tabular}
\end{center}
\end{table*}
The timescales as well as $e'$ and $a'$ are listed in \autoref{tab:tcoalesce}. We find that the timescale of SMBH coalescence primarily depends on the mass ratio between two SMBHs, $q$, while the dependencies on the orbital parameters and assumed eccentricity are subdominant. Importantly, $t\sub{coa}$ is much shorter than the Hubble time in  models with $q=0.01$ and 0.1, and for $q=1.0$, $t\sub{coa}$ is about 5~Gyr. Therefore mergers between NSCs hosting a SMBH in each centre are promising sites of GW emission and exciting targets for upcoming observations.

%%%%%%%%%%%%%%%%%%%%%%%%%%%%%%%%%%%%%%%%%%%%%%%%%%
%%%%%%%%%%%%%%%%%%%%%%%%%%%%%%%%%%%%%%%%%%%%%%%%%%
\section{Summary and Conclusions}
\label{sec:summary}

\begin{table}
\begin{center}
\caption{Relevant processes shrinking the orbit of SMBHs in each phase. DF, SH, OE and GW stand for dynamical friction, stellar hardening, Ouroboros Effect and GW emission, respectively. Type-M and -O represent the initial configuration, same as the simulation setup. 
\label{tab:rel_process}}
\begin{tabular}{ccc}
\hline 
Phase           & type-M   & type-O  \\
\hline 
Pre-binary      & DF+OE    & DF      \\
Combined effect & DF+SH+OE & DF+SH   \\
Hard binary     & SH       & SH      \\
GW emission     & GW(+SH)  & GW(+SH) \\
\hline
\end{tabular}
\end{center}
\end{table}
The coalescence of SMBHs is one of the most interesting targets for upcoming GW observations. In this paper, we investigate a possible path to accelerate the coalescence of SMBHs due to the presence of host NSCs, bypassing the final parsec problem. We find that an interplay of traditional dynamical friction, stellar hardening and an extra deceleration force -- that we term the `Ouroboros Effect' -- play a role to decrease the SMBHB's orbit, allowing it to coalesce in less than a Hubble time. This effect is a result of the tidal disruption of the NSCs surrounding the SMBHs, a process which exerts a braking force onto the SMBHB. Because of the scale-free nature of gravity, the Ouroboros Effect must work not only on the NSC scale but also on larger galactic scales. Interestingly, rapid orbital decay in the first few periods was also reported to occur in mergers between star clusters \citep{Arca-Sedda2018} and between galactic nuclei \citep{Khan2016,Khan2018a}. It could plausibly be driven by the same mechanism. In \autoref{tab:rel_process}, we list the relevant processes in each phase that the SMBHs experience before their eventual coalescence. The extra deceleration force is most pronounced when the second SMBH is less massive, since dynamical friction becomes less effective in making a binary. When the mass ratio of the binary is close to unity, a hard binary is directly formed within a few periods of the initial merging orbital time.

The extra deceleration force is caused by stars that initially belong to the NSC of the other SMBH. Stars initially on the downstream side tend to lose angular momentum because they are pulled back by their own NSC core, including the SMBH, while stars initially on the upstream side gain angular momentum because they are pulled forward by the NSC core and downstream stars. The exchange of angular momentum and orbital energy lets the former fall towards the potential minimum of the merged system and the latter move away from it. Then stars initially on the downstream side strongly decelerate the other NSC. 

We find that the orbital decay of the SMBHBs is well modelled with a single power-law function during the hard binary phase, and the power-law slope, i.e. the efficiency of the orbital decay, mainly depends on the mass ratio between the two SMBHs. The decay slope we found is shallower (i.e. slower orbital decay) than that predicted by the theoretical model developed by previous studies. Note that the density and velocity structure of the merged system would be different from those assumed in the previous studies and higher resolution simulations are desirable to discuss the evolution of the SMBHB in the hard binary phase in more detail. Therefore we leave the direct comparison between simulations and the theoretical model for future studies. We estimated the timescale of SMBH coalescence based on the extrapolation of the power-law function and find that SMBHs with a mass ratio of 1:10 or 1:100 would emit GWs and coalesce within $\sim 100$\,Myr from the beginning of the NSC merger while for the equal-mass case the total time is longer, 5~Gyr, but still less than the age of the Universe. 

While some more factors, e.g. galaxy merger rate, formation rate of NSCs, timescale of NSC approach after a galaxy merger and fraction of nucleated galaxies, must be taken into account to make predictions for observations, our estimation would be a positive implication for the future GW observations of low frequencies, such as the LISA, and point to the importance of nucleated galaxies in the low-mass regime. 

Our investigations also open other avenues of exploration. For instance, if stars in NSCs are mass segregated and heavier stars tend to sink in the centre of the cluster, the efficiency of stellar hardening may be enhanced by having heavy stars tightly bound to the central SMBHs. Another line of research relates to hyper-velocity stars: ejection of stars during the hard binary phase from the large supply of the merged NSC would be a signature of this process and can explain the detection of hyper-velocity stars from external galaxies \citep{Erkal2019}.

%%%%%%%%%%%%%%%%%%%%%%%%%%%%%%%%%%%%%%%%%%%%%%%%%%
%%%%%%%%%%%%%%%%%%%%%%%%%%%%%%%%%%%%%%%%%%%%%%%%%%
\section*{Acknowledgements}
We are grateful to the developers of {\sc Nbody6++gpu} for making their code publicly available, and thank Saavik Ford, Alessandra Mastrobuono-Battisti, Nadine Neumayer, Mathias Schultheis and Scott Tremaine for fruitful discussions. We acknowledge helpful comments from the anonymous referee and Manuel Arca-Sedda. GO and OH acknowledge funding from the European Research Council (ERC) under the European Union's Horizon 2020 research and innovation programme (grant agreement No. 679145, project `COSMO-SIMS'). The Flatiron Institute is supported by the Simons Foundation.



\bibliographystyle{mnras}
\bibliography{bhb}


%%%%%%%%%%%%%%%%%%%%%%%%%%%%%%%%%%%%%%%%%%%%%%%%%%
%%%%%%%%%%%%%%%%% APPENDICES %%%%%%%%%%%%%%%%%%%%%
%%%%%%%%%%%%%%%%%%%%%%%%%%%%%%%%%%%%%%%%%%%%%%%%%%
\appendix

%%%%%%%%%%%%%%%%%%%%%%%%%%%%%%%%%%%%%%%
\section{Stability of the NSC model in isolation}
\label{app:stability}

\begin{figure}
\begin{center}
\includegraphics[width=0.45\textwidth]{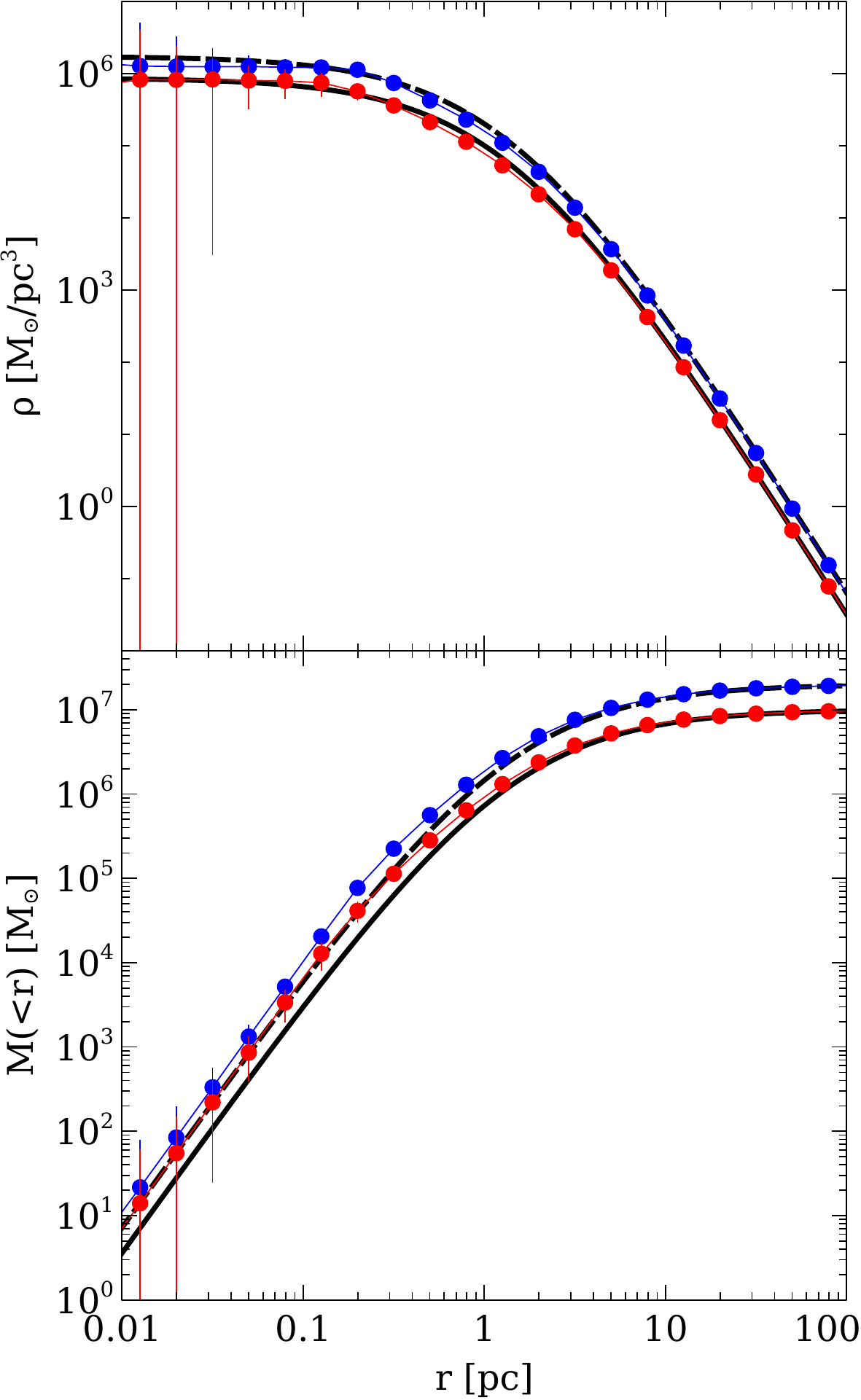}
\end{center}
\caption{
Radial profiles of stellar mass density ({\it upper}) and enclosed stellar mass ({\it lower}) of the NSC models. Models with a stellar mass of $10^7$ (red) and $2 \times 10^7$\,\msun{} (blue) evolve in isolation for 20\,Myr with a central SMBH with a mass of $10^6$\,\msun{}. Black solid and dashed lines show the analytical expression of the initial configuration of models with a stellar mass of $10^7$ and $2 \times 10^7$\,\msun{}, respectively. The origin is taken to be the cluster centre to draw the density profile of the simulated NSCs. In isolation, the NSC models are in a state of dynamical equilibrium. 
\label{fig:stab}}
\end{figure}
We test the stability of our NSC models hosting a central SMBH of $10^6$\,\msun{} by following the dynamical evolution of the systems in isolation. \autoref{fig:stab} depicts the radial density (upper) and mass profiles (lower) of the NSC models and shows that the NSCs reasonably keep their initial configuration at least for 20\,Myr which corresponds to $\sim 300 \ (370) \tau$ for the model with a stellar mass of $M\sub{nsc,tot} = 10^7 \ (2 \times 10^7)$\,\msun{}. The large scatter in the central region ($r \la 0.03$\,pc) where the enclosed mass falls below the mass resolution ($\sim 150$\,\msun{}) is due to Poisson noise. While this seems to imply that our results are not reliable in this radial range because of a lack of particles, the main results are insensitive to the number of particle (see \autoref{fig:orbit_vary_q}). We also observe that the SMBH settles in the centre of the NSC in the simulations of the NSC models, as expected.

%%%%%%%%%%%%%%%%%%%%%%%%%%%%%%%%%%%%%%%
\section{Accuracy parameters}
\label{app:force}

\begin{figure}
\begin{center}
\includegraphics[width=0.45\textwidth]{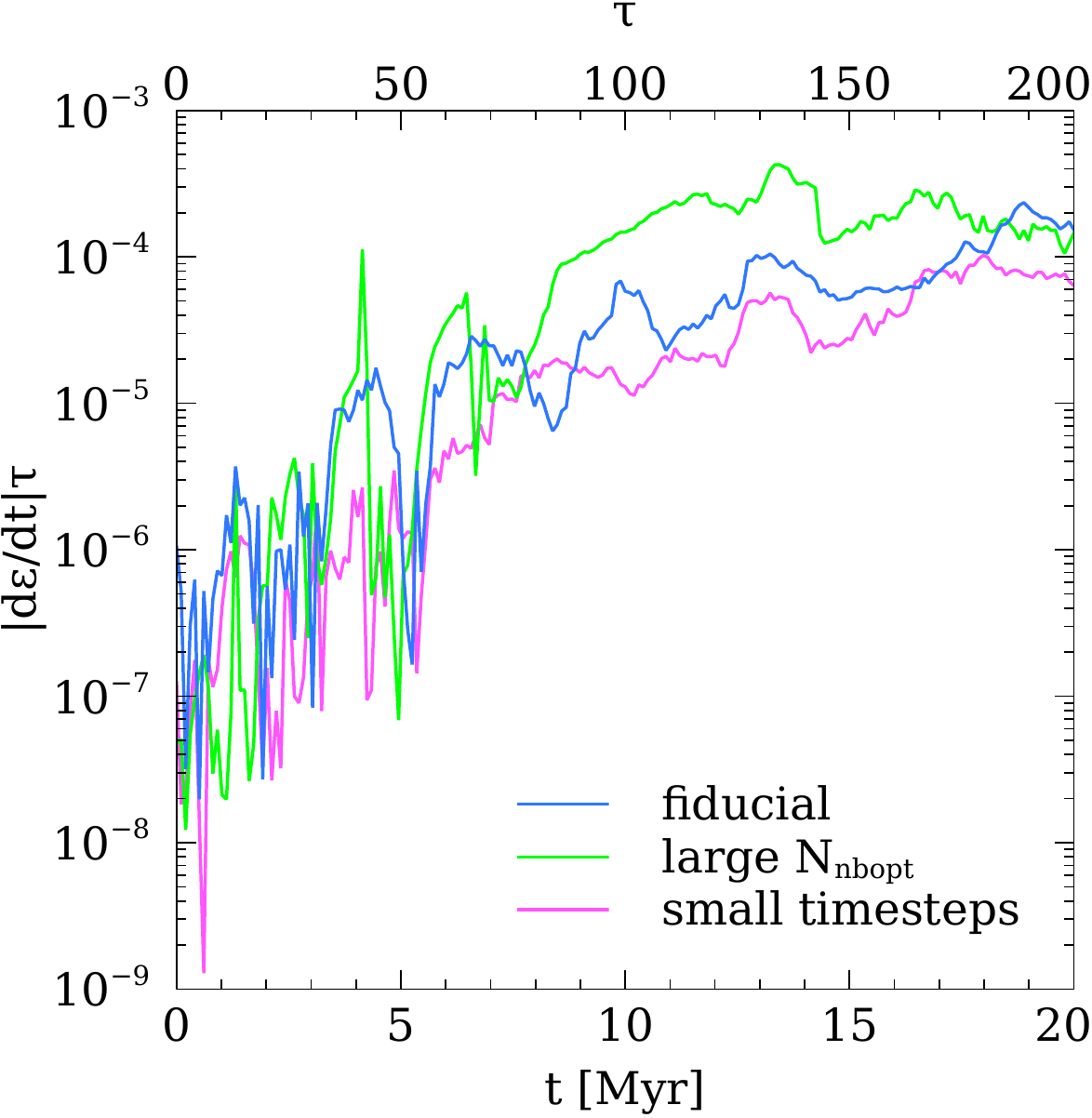}
\end{center}
\caption{
Relative energy error, $\mathcal{E}$, originated in one $N$-body time unit, $\tau$, in the type-M model of $q=0.1, d\sub{i}=20$\,pc and $\eta=1.0$. Blue line presents the results of simulations with the fiducial parameter set ($N\sub{nbopt}=64$, $\eta\sub{r}=\eta\sub{i}=0.005$ and $\eta\sub{u}=0.05$). The simulations of green and pink lines adopt the larger $N\sub{nbopt}=550$ ($\eta\sub{r}$, $\eta\sub{i}$ and $\eta\sub{u}$ remain as the fiducial values) and smaller timesteps ($\eta\sub{r}=\eta\sub{i}=0.0035$ and $\eta\sub{u}=0.035$) while $N\sub{nbopt}$ remains as the fiducial value. The level of energy conservation is comparable in the simulations.
\label{fig:ene_acc_param}}
\end{figure}

\begin{figure}
\begin{center}
\includegraphics[width=0.45\textwidth]{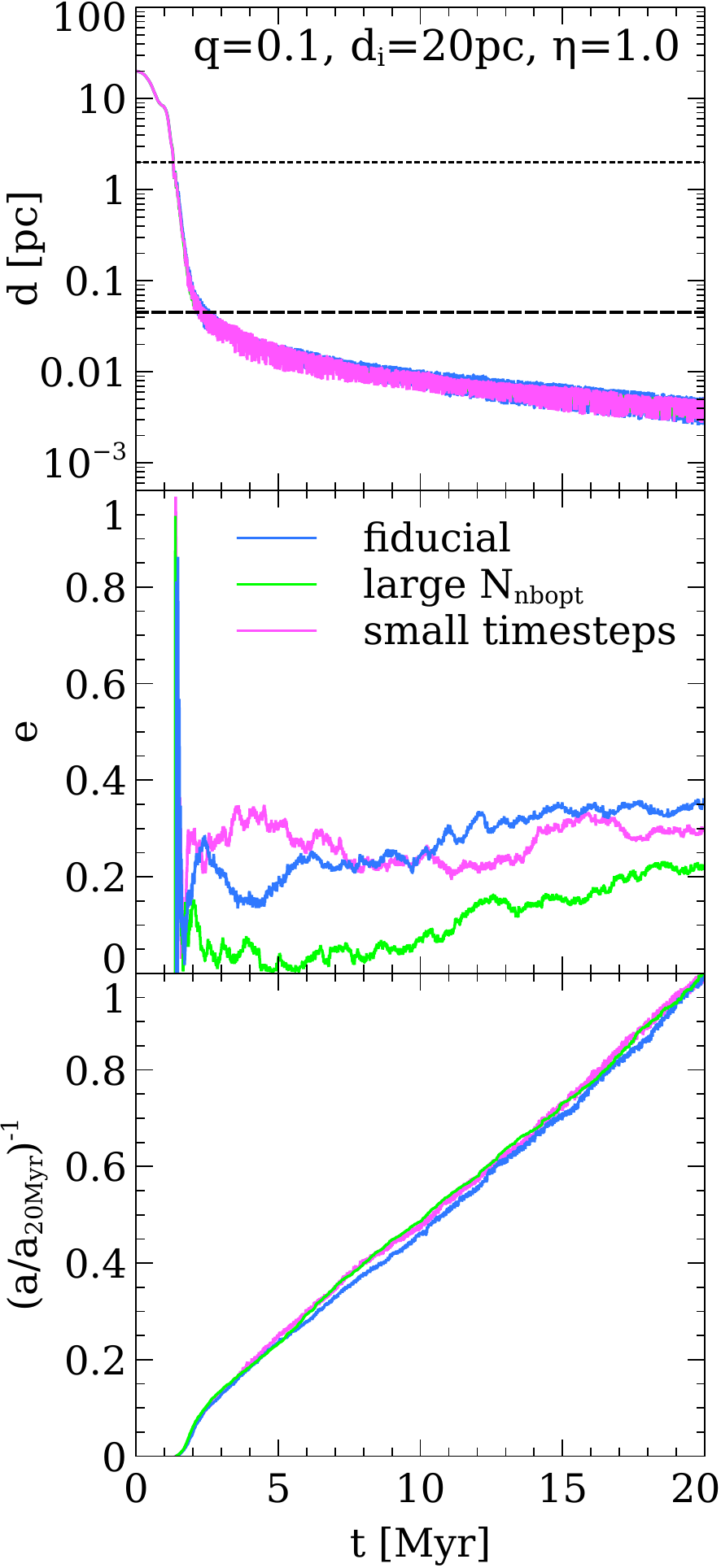}
\end{center}
\caption{
Evolution of the relative orbit of the SMBHs in the type-M model of $q=0.1, d\sub{i}=20$\,pc and $\eta=1.0$. Blue line presents the results of simulations with the fiducial parameter set ($N\sub{nbopt}=64$, $\eta\sub{r}=\eta\sub{i}=0.005$ and $\eta\sub{u}=0.05$). The simulations of green and pink lines adopt the larger $N\sub{nbopt}=550$ ($\eta\sub{r}$, $\eta\sub{i}$ and $\eta\sub{u}$ remain as the fiducial values) and smaller timesteps ($\eta\sub{r}=\eta\sub{i}=0.0035$ and $\eta\sub{u}=0.035$) while $N\sub{nbopt}$ remains as the fiducial value. 
({\it Top}) Separation between two SMBHs. Dotted and dashed horizontal lines are the gravitational influence radius of the primary SMBH, $d\sub{b}$, and hard binary separation, $d\sub{hb}$, estimated with the mass profiles of the merged system. The mass profiles are derived by stacking and averaging the snapshots in the simulation of the fiducial parameter set. 
({\it Middle}) Eccentricity of the orbit of the SMBHB, $e$. 
({\it Bottom}) Inverse semi-major axis of the SMBHB, $1/a$, normalized by that at $t=20$\,Myr. In computing $e$ and $a$, only the two SMBHs are taken into account, i.e. stellar particles are neglected. The evolution of the SMBHB is insensitive to the choice of parameters controlling the accuracy of orbital integration and $N\sub{nbopt}$. 
\label{fig:orbit_acc_param}}
\end{figure}

Here we test the validity of the numerical parameters adopted in our main simulations. 
\autoref{fig:ene_acc_param} compares the relative energy error, $\mathcal{E}$, in three simulations varying the parameters and finds the energy conservation of the comparable level. \autoref{fig:orbit_acc_param} shows the orbital evolution of the SMBHB and we find that our fiducial parameter set (blue) brings the compatible results with those in the simulation of the larger $N\sub{nbopt}$ (green) and smaller timesteps (pink) while the increase of eccentricity is delayed in them. This does not change our conclusion since the SMBHB coalescence timescale hardly depends on $e$ (\autoref{tab:tcoalesce}). Therefore our main simulations have the sufficient accuracy to study the dynamic evolution of the SMBHBs.

%%%%%%%%%%%%%%%%%%%%%%%%%%%%%%%%%%%%%%%
\section{Collisionless simulations}
\label{app:collisionless}
\begin{figure}
\begin{center}
\includegraphics[width=0.45\textwidth]{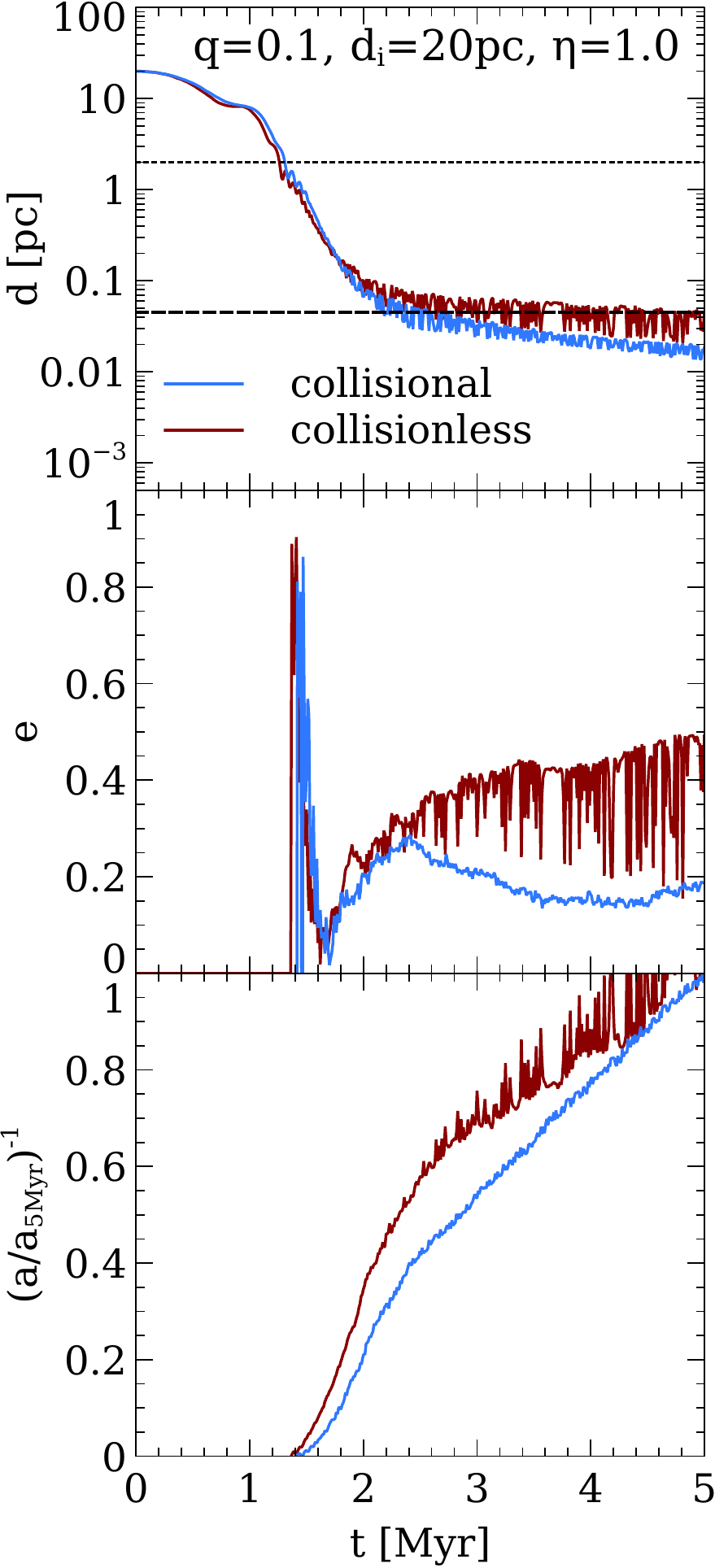}
\end{center}
\caption{
Evolution of the relative orbit of the SMBHs in the type-M model of $q=0.1, d\sub{i}=20$\,pc and $\eta=1.0$. Blue and brown lines present the results of simulations using $N$-body codes for collisional and collisionless dynamics, respectively. The same initial condition that consists of two SMBH particles and 131,072 stellar particles (i.e. our fiducial resolution) has been adopted. 
({\it Top}) Separation between two SMBHs. Dotted and dashed horizontal lines are the gravitational influence radius of the primary SMBH, $d\sub{b}$, and hard binary separation, $d\sub{hb}$, estimated with the mass profiles of the merged system. The mass profiles are derived by stacking and averaging the snapshots in the collisional simulation. 
({\it Middle}) Eccentricity of the orbit of the SMBHB, $e$. 
({\it Bottom}) Inverse semi-major axis of the SMBHB, $1/a$, normalized by that at $t=5$\,Myr. In computing $e$ and $a$, only the two SMBHs are taken into account, i.e. stellar particles are neglected. 
As expected, the two simulation codes show a good agreement in the collisionless regime, i.e. before entering the hard binary phase ($t \la 2$\,Myr). 
\label{fig:vs_tree}}
\end{figure}
The drag forces that drive the rapid orbital decay of the SMBHs in the first few Myrs of the NSC merger, i.e. the Ouroboros Effect and dynamical friction, are collisionless processes. Since the number of stellar particles employed in the simulations ($\sim 10^5$) is smaller than that of stars in real NSCs ($\sim 10^7$), collisionality in the simulated systems is higher than in real NSCs. While the orbital evolution of the SMBHB is insensitive to the number of stellar particles in collisional simulations (\autoref{fig:orbit_vary_q}), we additionally perform a collisionless simulation to address the importance of collisionality in this study. A treecode \citep{Barnes1986} accelerated with Graphics Processing Units \citep{Ogiya2013} is used for this collisionless simulation. To ensure a collisioless nature of the system, the gravitational potential field of particles is softened by introducing the force softening, $\epsilon$, that effectively sets the spatial resolution of the simulations. We employ a Plummer force softening \citep{Plummer1911} with a softening length $\epsilon = 0.01$\,pc, and a cell opening criteria following \cite{Springel2005} with the parameter controlling the force accuracy set to $\alpha=0.01$. The second order Leapfrog scheme with the variable timestep \citep{Power2003} is used for orbit integration.

\autoref{fig:vs_tree} compares the orbital evolution of SMBHs in collisonal (blue) and collisionless (brown) simulations. The two simulations show an excellent qualitative agreement before the SMBHs form a hard binary ($d \ga d\sub{hb}$) and collisional stellar hardening sets in. This result verifies that the drag forces are indeed collisionless processes. Also, \autoref{fig:orbit_vary_q} indicates that our main results are insensitive to the possible effect of artificially high collisionality due to the small number of particles. The top panel of \autoref{fig:vs_tree} also shows that after the SMBHs form a hard binary, the orbital decay is slower in the collisionless simulation. This is mainly due to the softened gravitational potential field. The separation between the SMBHs is comparable to the force softening, so that the subsequent dynamical evolution in the collisionless simulation is unresolved. On the other hand, the collisional simulation continues to follow the dynamical evolution of the merged system in the framework of the pure Newtonian dynamics since the force softening is not included. Due to the lack of accuracy in force computation and orbit integration, the orbital evolution of the SMBHB in the collisional regime is noisier in the collisionless simulation than in the collisional simulation, as shown in the middle and bottom panels. Therefore, a colissional simulation code is indeed more suited for the purpose of this study even though, of course, there are never enough particles in an $N$-body simulation.

%%%%%%%%%%%%%%%%%%%%%%%%%%%%%%%%%%%%%%%%%%%%%%%%%%
%%%%%%%%%%%%%%%%%%%%%%%%%%%%%%%%%%%%%%%%%%%%%%%%%%
% Don't change these lines
\bsp	% typesetting comment
\label{lastpage}
\end{document}